\documentclass[a4paper,11pt]{article}
\pdfoutput=1 

\usepackage{jheppub} 

\usepackage[T1]{fontenc} 

\usepackage[dvipsnames]{xcolor}
\usepackage{pdfpages}

\usepackage{subcaption}

\newcommand{\nn}{\nonumber\\ }

\def\dd{\text{d}}

\def\med{\text{med}}

\newcommand{\order}[1]{{\cal O}\left(#1\right)}
\newcommand{\as}{\alpha_s}

\def\cP{\mathcal{P}}
\def\k{\boldsymbol{k}}

\DeclareMathOperator\erf{erf}

\newcounter{RSQ}

\preprint{CERN-TH-2023-243
}
\title{Advancing the understanding of energy-energy correlators in heavy-ion collisions}

\author[a]{Jo\~ao Barata,}
\author[b]{Paul Caucal,}
\author[c]{Alba Soto-Ontoso,}
\author[a]{and Robert Szafron}
\affiliation[a]{Physics Department, Brookhaven National Laboratory, Upton, NY 11973, USA}
\affiliation[b]{SUBATECH UMR 6457 (IMT Atlantique, Université de Nantes,
IN2P3/CNRS), 4 rue Alfred Kastler,44307 Nantes, France}
\affiliation[c]{CERN, Theoretical Physics Department, CH-1211 Geneva 23, Switzerland}
\abstract{
We investigate the collinear limit of the energy-energy correlator (EEC) in a heavy-ion context. First, we revisit the leading-logarithmic (LL) resummation of this observable in vacuum following a \textit{diagrammatic} approach. We argue that this route allows to naturally incorporate medium-induced effects into the all-orders structure systematically. As an example, we show how the phase-space constraints imposed by the medium on vacuum-like emissions can be incorporated into the LL result by modifying the anomalous dimensions. On the fixed-order side, we calculate the $\mathcal{O}(\alpha_s)$ expansion of the in-medium EEC for a $\gamma\to q\bar q$ splitting using, for the first time, the exact matrix element. When comparing this result to previously used approximations in the literature, we find up to $\mathcal{O}(1)$ deviations in the regime of interest for jet quenching signatures. Energy loss effects are also quantified and further suppress the EEC at large angles. These semi-analytic studies are complemented with a phenomenological study using the jet quenching Monte Carlo \texttt{JetMed}. Finally, we argue that the imprint of medium-induced effects in energy-energy correlators can be enhanced by using an alternative definition that takes as input Lund primary declusterings instead of particles. 
}

\begin{document}
\maketitle

\section{Introduction}
Jet substructure observables have become an essential ingredient in the search for quark-gluon plasma (QGP) signatures in heavy-ion collisions in recent years. Their inherent sensitivity to disparate scales fits very well with the multi-scale nature of jet evolution in the QGP. From a theoretical point of view, they also offer a unique opportunity to produce first principles predictions using resummation techniques in perturbative QCD. This semi-analytic approach is very well established in proton-proton collisions, and the heavy-ion community has frequently benefited from these developments. As an example, the first set of jet substructure measurements in heavy-ion collisions (and related calculations) focused on Soft Drop grooming~\cite{Larkoski:2014wba,CMS:2017qlm,Chien:2016led,Mehtar-Tani:2016aco}. However, it quickly became apparent that porting proton-proton-based observables to the complex environment of heavy-ion collisions had some limitations, mainly due to the high hadronic multiplicity. This has triggered the development of new experimental and theoretical techniques with a heavy-ion rationale. The advances in the field of jet substructure in heavy-ions have been so rapid that experiments such as ALICE or STAR (traditionally heavy-ion focused) have pioneered measurements of such class of observables even in proton-proton collisions~\cite{ALICE:2021aqk,ALICE:2022hyz,Song:2023sxb}. The most recent example of a jet substructure observable that has attracted ample attention are energy flow correlations~\cite{Basham:1978zq}, which are being actively measured in several experimental setups~\cite{alice-talk, star-talk}.

Formally, energy-flow correlators are defined in terms of light-ray energy flow operators along direction $\vec n$~\cite{Sveshnikov:1995vi,Tkachov:1995kk,Korchemsky:1999kt,Bauer:2008dt,Hofman:2008ar,Belitsky:2013xxa,Kravchuk:2018htv}
\begin{equation}
\mathcal{E}(\vec n) = \lim_{r\to\infty} \int_0^\infty \dd t\, r^2 n^i T_{0i}(t,r \, \vec n)  \, ,
\end{equation}
with the stress-energy tensor $T_{\mu\nu}$. This paper focuses on the simplest projection of the two-point correlator (EEC), where the azimuthal dependence is integrated out. This results in the standard definition of the EEC:
\begin{align}
\label{eq:eec-definition}
\frac{\dd\Sigma^{(n)}}{\dd \chi} &= \int_{\vec n_1,\vec n_2} \frac{\langle \mathcal{E}^n(\vec n_1) \mathcal{E}^n(\vec n_2) \rangle}{Q^2} \delta(\vec{n_1}\cdot \vec{n_2}-\cos\chi)  \nn 
&=\sum_{\lbrace i, j\rbrace} \int_{\vec n_i,\vec n_j} \frac{E_i^n E_j^n}{Q^{2n}} \frac{\dd\sigma}{\dd \vec n_i \dd \vec n_j} 
\delta(\vec{n}_i\cdot \vec{n}_j-\cos\chi) \, ,
\end{align}
where $n$ is an integer number that we set to $1$ by default, dropping the superscript notation in the EEC, and we explore its variation in Sec.~\ref{sec:LEEC}. In the second line, we have introduced the inclusive cross-section for producing two particles with energies $E_{i,j}$ along the directions $\vec n_{i,j}$, and $Q$ is the virtuality of the hard process. The sum over $\{i,j\}$ includes all unordered particle pairs. 

The first studies of EECs date back to the late 1970's~\cite{PhysRevD.17.2298, Basham:1979gh,Basham:1978zq} in the context of testing and verifying QCD properties. Fixed-order calculations presented in these works were used to extract the strong coupling constant, $\alpha_s$, in $e^+e^-$ colliders~\cite{OPAL:1990reb,DELPHI:1990sof, SLD:1994idb}. In the collinear and back-to-back limits, the EEC receives large logarithmically enhanced corrections that must be resummed to all orders in the strong coupling constant. Substantial theoretical effort has been devoted to significantly improve the analytic description of this observable both in QCD~\cite{deFlorian:2004mp,Dokshitzer:1999sh,Ebert:2020sfi,Tulipant:2017ybb,DelDuca:2016ily,Dixon:2019uzg,Kardos:2018kqj,Lee:2022ige,Chen:2023zlx} and $\mathcal{N}=4$ SYM theory~\cite{Hofman:2008ar,Belitsky:2013xxa,Belitsky:2013ofa,Belitsky:2013bja,Henn:2019gkr}. Despite its apparent simplicity, the EEC 
has a wide range of applications in the high energy context, including top mass extraction~\cite{Holguin:2022epo}, gluon saturation~\cite{Liu:2023aqb}, and the confining transition~\cite{Komiske:2022enw}. One of the most recent applications of the EECs has been the extraction of $\alpha_s$ at the LHC~\cite{CMS:2023wcp}.

In the context of heavy-ion collisions, the EECs could potentially address a series of questions related to the interaction of a jet with the medium, such as: (i) is the critical temperature of the QGP imprinted in the jet fragmentation?, (ii) what is the angular resolution of the medium?, (iii) can we experimentally disentangle medium response from transverse momentum broadening? Furthermore, the energy weighting entering Eq.~\eqref{eq:eec-definition} could help suppress the copious soft contamination arising from the underlying event. During the last year, there has been a series of studies discussing the capability of the EEC to resolve the scales of the QGP~\cite{Andres:2022ovj,Andres:2023xwr,Barata:2023zqg,Andres:2023ymw,Yang:2023dwc}. 
The analytic calculations in the medium are typically performed at the leading order using certain approximations for the medium modified jet cross-section. Refs.~\cite{Andres:2022ovj,Andres:2023xwr,Barata:2023zqg,Andres:2023ymw} used a semi-classical approximation in the BDMPS-Z formalism that was derived in Ref.~\cite{Dominguez:2019ges}. In turn, Ref.~\cite{Yang:2023dwc} provides a leading-order calculation using the higher-twist formalism~\cite{Wang:2001ifa}. Indistinctly of the specific approximation used for the leading-order matrix element, all these works predict an enhancement at large angles due to medium-induced modifications. However, once other effects such as medium response or energy loss are taken into account either analytically~\cite{Barata:2023vnl} or by means of realistic Monte Carlo (MC) simulations~\cite{Wang:2001ifa}, the interpretation of the EEC becomes less transparent. Most of these processes take place at commensurate scales, and therefore, resolving each of them independently is a remarkably complicated task, as has been observed in many other jet substructure observables. The intense experimental activity around the EECs in the heavy-ion community calls for a focused theoretical effort to guarantee a solid interpretation of the upcoming experimental data. 

This paper takes a step towards increasing the precision of the theoretical tools required to describe the EEC in heavy-ion collisions and provides a careful assessment of the size of different effects. We do so by combining semi-analytic calculations with parton shower simulations. In Sec.~\ref{sec:resum}, we re-derive the leading logarithmic resummation of the EEC following a diagrammatic approach in vacuum. This alternative way of performing the calculation aims to incorporate the medium modifications, as shown in Sec.~\ref{sec:medium}. Throughout this section, we explore three different sources of medium modifications to the EEC. First, we study in Sec.~\ref{sec:med_phase_space} the phase-space modification of vacuum emissions in the presence of a medium. In Sec.~\ref{sec:med-split-function} we discuss three different calculations of the medium modification factor for two different partonic channels: exact result in the multiple-soft scattering approximation~\cite{Isaksen:2023nlr}, semi-classical approximation~\cite{Dominguez:2019ges} and BDMPS-Z result~\cite{Baier:1996kr,Zakharov:1996fv}. A simple energy loss model is presented in Sec.~\ref{sec:e-loss}. Results for the EEC at leading-order can be found in Sec.~\ref{sec:numerics_EEC}. We find that an exact treatment of the leading order cross-section strongly reduces the enhancement at large angles observed in previous works. After including energy loss, the significance of the signal at large angles is even further reduced. Besides these analytic estimates, we also perform a MC study of this observable within the {\tt JetMed} framework as shown in Sec.~\ref{sec:results-jetmed}.  We end up by proposing a new definition of the EEC in terms of Lund subjets~\cite{Dreyer:2018nbf} in Sec.~\ref{sec:LEEC} and summarising our results in Sec.~\ref{sec:conclusions}.

\section{Revisiting the EEC resummation in a diagrammatic approach}
\label{sec:resum}
In this first section, we present a derivation of the EEC cumulative distribution at leading logarithmic (LL) accuracy in the collinear limit. In contrast with previous derivations making explicit use of the light-ray OPE for the energy flow operators~\cite{Chen:2020vvp,Chen:2021gdk,Chen:2023zzh}, or Effective Field Theory (EFT) techniques~\cite{Dixon:2019uzg},
here we obtain the final distribution by summing infinite set of relevant Feynman diagrams.\footnote{For a review of the methods to be employed see for example~\cite{Marzani:2019hun} and references therein.} Although the final expressions for the cumulative distribution agree using any of these methods, the diagrammatic approach (also used in \cite{Konishi:1979cb,Richards:1982te}) is arguably more convenient in the jet quenching context, where a QCD medium introduces multiple emergent scales and induces modifications to the jet fragmentation pattern. A complete operatorial level or EFT description for the case of jet evolution in dense QCD matter is not yet available (see~\cite{Vaidya:2020lih,Ovanesyan:2011xy} for related efforts). The description of these effects in the language of light-ray OPE is also not trivial since it requires going beyond the conformal limit of the theory. We note that in real-world QCD, the conformal symmetry is broken by multiple effects, some present in vacuum (running coupling or quark masses) and others related to medium scales. In contrast, medium modifications to jet evolution are relatively well understood using standard diagrammatic methods, see e.g.~\cite{Mehtar-Tani:2013pia,Blaizot:2015lma}, and form the backbone of many jet quenching phenomenological models. In what follows, we first present a complete derivation of the cumulative distribution assuming a quark-initiated jet and keeping only the Abelian contributions (i.e.,, equivalent to QED when ignoring photon branching). We then point out how this generalizes to the well-known QCD result.

Consider a jet with a radius $R$ and total energy $Q$.\footnote{We note that we use energy and transverse momentum interchangeably throughout this section. In particular, we shall take $Q=p_t$, with $p_t$ equal to the total transverse momentum of the jet in laboratory coordinates.} It is convenient to rewrite Eq.~\eqref{eq:eec-definition} as
\begin{align}
\label{eq:EEC_def}
\frac{\dd \Sigma}{\dd \chi} = \frac{1}{\sigma}\sum_{ \{i,j\}\in {\rm jet} }\int_0^1 \dd z \frac{\dd\sigma}{\dd \theta_{ij} \dd z} z(1-z) \delta\left(\chi-\frac{\theta_{ij}}{R}\right)\, ,
\end{align}
where $z$ denotes the energy fraction $E_i/Q$ of the softest parton in a particle pair separated by a distance $\theta_{ij}<R$, and we have simplified the measurement function to the collinear form. We will focus on the cumulative distribution $\Sigma(\chi)$.

As mentioned above, we consider only the Abelian channel of a quark-initiated jet, where there are only two relevant splitting functions, namely 
\begin{align}
\label{eq:splitting_functions}
  &P_{qq}(z) = C_F\frac{1+z^2}{1-z}\, , \quad P_{gq}(z) = C_F\frac{1+(1-z)^2}{z} \, .
\end{align}
In what follows, we use the fact that $P_{gq}(z)=P_{qq}(1-z)$ to write all the contributions in terms of $P_{gq}$ and, therefore, $z\to 0$ corresponds to the soft gluon limit. The one-gluon emission matrix-element squared, at fixed coupling, is thus given by
\begin{align}\label{eq:splitting_fucntions_2}
\dd \mathcal{P}^{\rm vac} =\bar\alpha P_{gq}(z) \dd z \frac{\dd \theta}{\theta}\,,
\end{align}
where $\bar\alpha=\as/\pi$ and we have again exploited the fact that we work in the collinear limit at LL accuracy to keep only the logarithmically divergent piece on the angular dependence. Note that all angles are normalized by the jet radius. 

\begin{figure}[h!]
    \centering
    \includegraphics[width=.75\textwidth]{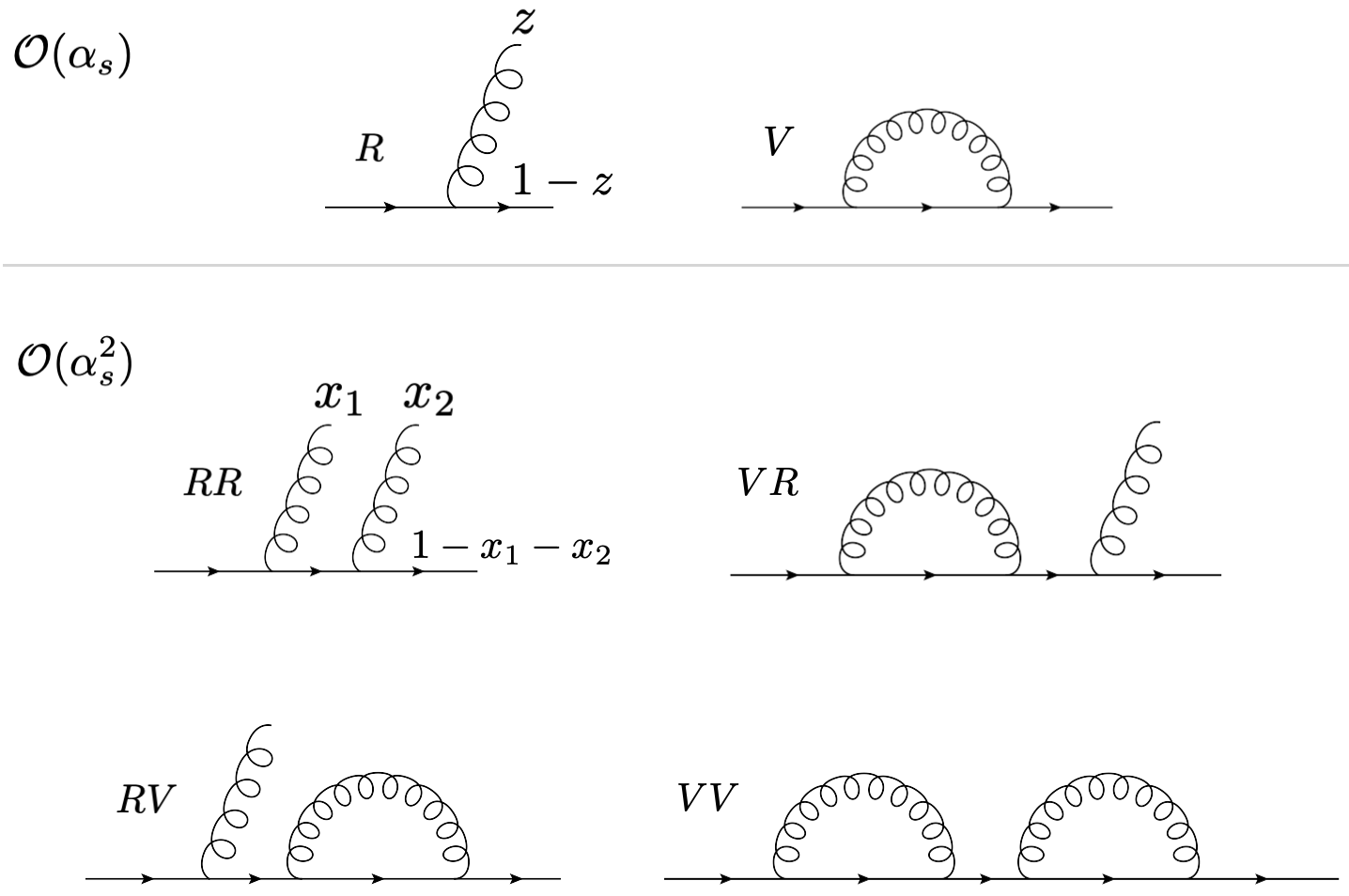}
    \caption{Leading order diagrams contributing to the quark EEC in the Abelian channel. Conventions used in the main text for the energy fraction for the real diagrams are shown.}
    \label{fig:diagrams_RV}
\end{figure}
 At first order in the strong coupling constant, i.e., $\mathcal{O}(\alpha_s)$, there are two contributions to the observable: one with a real (R) gluon emission and a diagram with a virtual (V) correction, see Fig.~\ref{fig:diagrams_RV}. The contribution of the real gluon emission to $\Sigma$ is:
\begin{align}\label{eq:as-real}
    \Sigma^{\rm R}(\chi)\Big\vert_{\order{\as}} &=\bar \alpha \int_0^1 \frac{\dd\theta}{\theta}  \int_0^1  \dd z \, P_{gq}(z) \Big[z^2 \Theta(\chi) +(1-z)^2 \Theta(\chi) +2z(1-z)\Theta(\chi-\theta)\Big] \, .
\end{align}
The first two terms in the previous equation correspond to measuring the EEC on the same parton (i.e., $i=j$ term in Eq.~\eqref{eq:EEC_def}), while the last piece accounts for twice the contribution coming from the resolved EEC (i.e., $i\neq j$ term in Eq.~\eqref{eq:EEC_def}). The virtual contribution can be written as
\begin{align}
\label{eq:as-virtual}
   \Sigma^{\rm V}(\chi)\Big\vert_{\order{\as}} = - \bar \alpha \int_0^1 \frac{\dd\theta}{\theta}  \int_0^1  \dd z \, P_{gq}(z)  \Theta(\chi) \; ,
\end{align}
where the overall minus sign ensures the exact cancellation of the divergences between the real and virtual diagrams. As a result, we find that the $\mathcal{O}(\alpha_s)$ cumulative distribution reads 
\begin{align}
\label{eq:eec-LO}
  \Sigma(\chi)\Big\vert_{\order{\as}} &=\bar \alpha  \int_0^1 \frac{\dd \theta}{\theta} \dd z P_{gq}(z)2z(1-z) \left[ \Theta(\chi-\theta) -\Theta(\chi)\right] \nn 
  &= -2 \bar\alpha \ln \frac{1}{\chi} \; \int_0^1  \dd z \, P_{gq}(z)z(1-z) \;,
\end{align}
where the remaining integral is finite, and we postpone its explicit calculation until the end of this section.

At $\order{\as^2}$, and again keeping only the Abelian contributions, we find 4 distinct diagrams as depicted in Fig.~\ref{fig:diagrams_RV}. We introduce the energy fraction of a real emission with respect to the jet initiating parton $x_i = E_i/Q$, which only coincides with the local energy fraction $z_i$ for the first emission. Energy degradation along the primary branch implies that for any other emission $x_i = \Pi_{j<i} (1-z_j)z_i$, such that energy conservation reads $\sum_{i=1}^n x_i = 1$ (with $n=3$ at this order in $\as$). We write the results in terms of three angles: $\theta_1\equiv\theta_{13}, \theta_2\equiv\theta_{23}$ and $\theta_{12}$, where $\theta_{12}^2 = \theta_1^2 + \theta_2^2 -2\theta_1\theta_2 \cos\phi_{12}$ and $\phi_{12}$ denotes the relative azimuthal angle between the gluons. At leading-logarithmic accuracy, we can further impose strong angular ordering between real emissions such that $\theta_1\gg\theta_2$ and $\theta_{12}\sim \theta_1$.

Taking into account all these considerations, the RR diagram gives
\begin{align}
\label{eq:sigma-RR}
\Sigma^{\rm{RR}}(\chi)\Big\vert_{\order{\as^2}} &=\bar\alpha^2 \int_0^1 \frac{\dd\theta_1}{\theta_1}\int_0^1 \frac{\dd\theta_2}{\theta_2} \int_0^{2\pi}\frac{\dd \phi_{12}}{2\pi}\int_0^1\dd x_1\,  P_{gq}(x_1) \int_0^{1-x_1} \frac{\dd x_2}{1-x_1} \,  P_{gq}\Big(\frac{x_2}{1-x_1}\Big) \nn
&\times \Big[2 x_1 x_3 \Theta(\chi-\theta_1) +  2 x_2 x_3 \Theta(\chi-\theta_2) + 2 x_1 x_2 \Theta(\chi-\theta_{12}) + x_1^2 + x_2^2 + x_3^2 \Big] \, ,
\end{align}
with $x_3=1-x_1-x_2=(1-z_1)(1-z_2)$ and where in the last three terms inside the square bracket there is an implicit $\Theta(\chi)$. Note that the previous expression also accounts for the corresponding Jacobian when transforming from $z_i$ to $x_i$. For the RV case, we have 
\begin{align}
\label{eq:sigma-RV}
\Sigma^{\rm{RV}}(\chi)\Big\vert_{\order{\as^2}} & = -\bar\alpha^2 \int_0^1 \frac{\dd\theta_1}{\theta_1}\, \int_0^1 \frac{\dd\theta_2}{\theta_2}\, \int_0^1\dd x_1 \, P_{gq}(x_1) \int_0^1 \dd x_2 \, P_{gq}(x_2) \nn
&\times \Big[2 x_1 x_3 \Theta(\chi-\theta_1) + x_1^2 + x_3^2 \Big] \, ,
\end{align}
where $x_3 = 1-x_1$ since the second gluon is virtual. The VR diagram has the same structure as Eq.~\eqref{eq:sigma-RV}. Finally, the VV contribution has a single term
\begin{align}
\Sigma^{\rm{VV}}(\chi)\Big\vert_{\order{\as^2}} & = \bar\alpha^2\int_0^1 \frac{\dd\theta_1}{\theta_1}\int_0^1 \frac{\dd\theta_2}{\theta_2}\int_0^1\dd x_1 \, P_{gq}(x_1) \int_0^1 \dd x_2 \, P_{gq}(x_2)\, .
\end{align}
The full $\mathcal{O}(\alpha_s^2)$ result is obtained by adding up all 4 contributions that are individually collinearly divergent. It is straightforward to show that the different diagrams cancel out exactly in all regions where double or single collinear poles could emerge, as expected. The only surviving contribution comes from the region where $\theta_{1}>\theta_{2}>\chi$, where the integrands combine to give:
\begin{align}
&\Big[ (x_1^2 + x_2^2+x_3^2)^{\rm{RR}} - (x_1^2+x_3^2)^{\rm{RV}} -(x_2^2+x_3^2)^{\rm{VR}} + (1)^{\rm{VV}}\Big]\Theta(\chi<\theta_1<1) \Theta(\chi<\theta_2<\theta_1)\, .
\end{align}
After some algebraic manipulation, we find
\begin{align}
\Sigma(\chi)\Big\vert_{\order{\as^2}} 
&= \frac{\bar \alpha^2}{2!} \ln^2\frac{1}{\chi} \int_0^1\dd z_1 P_{gq}(z_1) \int_0^1 \dd z_2 P_{gq}(z_2)  [2 (z_1-2) z_1 \,   (z_2-1) z_2] \, ,
\end{align}
where, again, the remaining integrals are finite. 

The previous calculation can be systematically extended to higher orders in $\as$ with the only surviving contribution coming from the region without poles order by order. In general, we find that the $\order{\as^k}$ contribution reads (here assuming $k>2$):
\begin{align}
\label{eq:sigma:allk}
  \Sigma(\chi)\Big\vert_{\order{\as^{k>2}}} &=\frac{2\, \bar \alpha^{k+1}}{(k+1)!} \ln^{k+1} \frac{1}{\chi} \left[ \prod_{l=1}^{k+1}\int_0^1 \dd z_l\, P_{gq}(z_l) (z_l-2) z_l\right] \frac{z_{k+1}-1}{z_{k+1}-2} \, .
\end{align}
The final distribution is then obtained by summing Eq.~\eqref{eq:sigma:allk} over $k$. To that end, we introduce the anomalous dimensions
\begin{align}
\label{eq:gamma_vac}
   \gamma_{ik}(j) = - \int_0^1 {\dd z} \, z^{j-1}  \hat P_{ik}(z)\, ,
\end{align}
where $j$ is an integer number, $(i,k)$ run over all possible flavors and $\hat P$ denotes the regularized splitting function, i.e., $\hat P$ corresponds to the LO kernels in Eq.~\eqref{eq:splitting_functions} evaluated using the plus-prescription. For the splitting functions of interest in our calculation, one has
\begin{align}
\label{eq:gammas-def}
   &\gamma_{qq}(j) =- C_F \left[\frac{3}{2} + \frac{1}{j(j+1)} - 2[\psi(j+1)+\gamma_E] \right]\, , \quad \gamma_{gq}(j) = -C_F \frac{2+j+j^2}{j^3-j}\, ,
\end{align}
where $\psi(j)\equiv \Gamma'(j)/\Gamma(j)$ is the di-gamma function and $\gamma_E$ is the Euler-Mascheroni constant. 
A direct calculation gives
\begin{align}\label{eq:resummed_gammas}
     \Sigma(\chi) &=\sum_{k=0}^{\infty}(-1)^{k+1}\frac{\bar \alpha^{k+1}}{(k+1)!} \ln^{k+1} \frac{1}{\chi} [\gamma_{qq}^{k+1}(3) + \gamma^k_{qq}(3)\gamma_{gq}(3)] \nn
     & = \left(-1+\chi^{\bar \alpha\gamma_{qq}(3)}\right)\frac{\gamma_{gq}(3)+\gamma_{qq}(3)}{\gamma_{qq}(3)}\, ,
\end{align}
which results in the following differential distribution 
\begin{align}
\label{eq:sigma-LL}
\frac{\dd\Sigma}{\dd \chi} &= \frac{\bar\alpha}{\chi^{1-\bar\alpha \gamma_{qq}(3)}}[\gamma_{gq}(3)+\gamma_{qq}(3)] =  \frac{\bar\alpha}{\chi^{1-\bar\alpha \gamma_{qq}(3)}}\int_0^1 \dd z\, z(1-z) \left[  P_{qq}(z)+P_{gq}(z)  \right] \, .
\end{align}
\begin{figure}
    \centering
    \includegraphics[width=.85\textwidth]{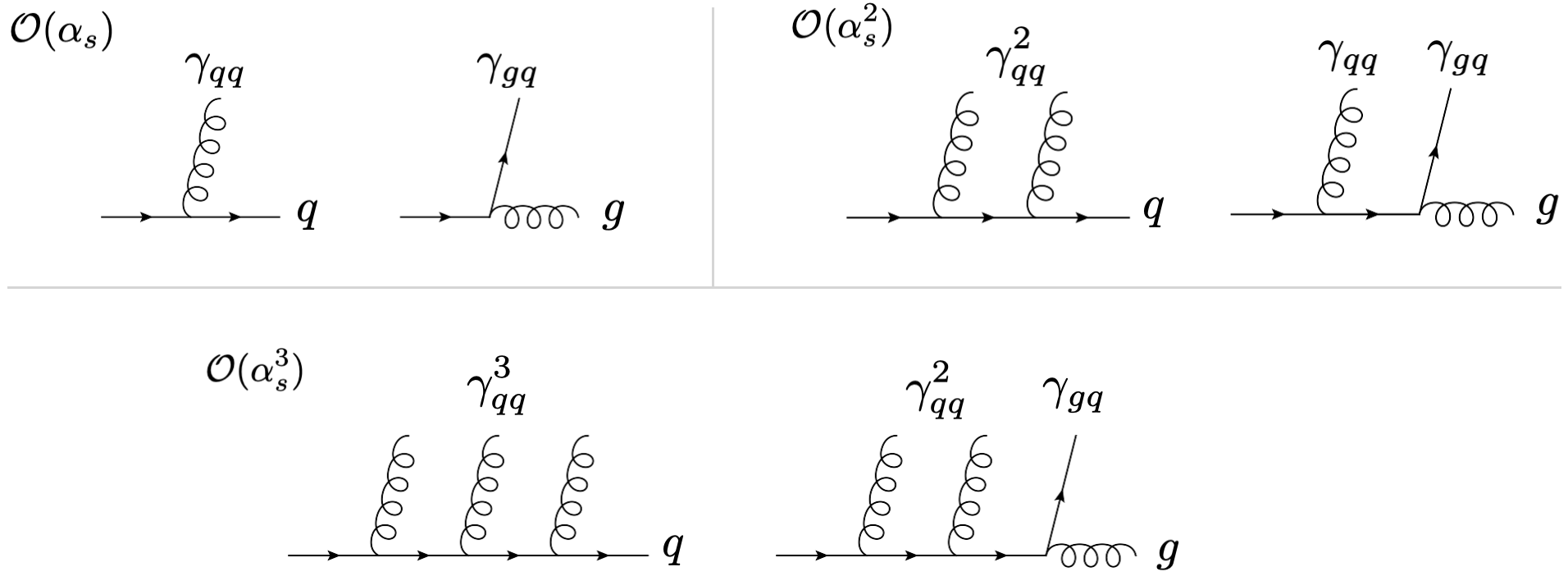}
    \caption{Diagrammatic representation of the resummation structure in terms of the anomalous dimensions. Note that since we only include the Abelian quark channel, flavor-changing branchings can only occur at the last step in the ladder, unlike the full QCD case. A similar representation can be drawn in QCD, agreeing with the structure of $\hat \gamma(j)$. }
    \label{fig:anomalous_dim}
\end{figure}
The form shown in the first line of Eq.~\eqref{eq:resummed_gammas} has a simple diagrammatic interpretation, illustrated in Fig.~\ref{fig:anomalous_dim}. It results from the repeated application of the anomalous dimension matrix, which, in the complete QCD case, reads 
\begin{align}
 \hat {\gamma}(j) = \begin{bmatrix}
\gamma_{qq}(j) & \gamma_{qg}(j) \\
\gamma_{gq}(j) & \gamma_{gg}(j)
\end{bmatrix}   \, ,
\end{align}
while in the Abelian quark sector one can set $\gamma_{gg}=\gamma_{qg}= 0$. Using this, a result analogous to the one shown in Eq.~\eqref{eq:sigma-LL} can be derived~\cite{Dixon:2019uzg}. 

\section{Medium modifications to the EEC}
\label{sec:medium}
This section aims to explore the medium modifications to the jet EEC. We discuss three possible sources: (i) phase-space constraints to vacuum-like splittings, (ii) medium modifications to the leading-order splitting function, and (iii) leading energy loss for resolved emissions. To this end, we assume a simple model for the QGP consisting of a finite slab of dense, static, isotropic and homogeneous matter.\footnote{For recent theoretical efforts towards describing jet evolution in more realistic matter, see e.g.~\cite{He:2020iow, Sadofyev:2021ohn, Antiporda:2021hpk, Hauksson:2021okc, Fu:2022idl, Barata:2022utc, Hauksson:2023tze, Boguslavski:2023alu, Barata:2023qds, Kuzmin:2023hko,Adhya:2021kws,Caucal:2020uic} and references therein.} The medium is assumed to have a longitudinal extension $L$. The jet-medium interactions are described using the multiple soft scattering approximation, detailed below. As a result, we neglect contributions from rare hard momentum exchanges between the jet and medium constituents, whose exclusion should not significantly change the final results at a qualitative level~\cite{Andres:2023xwr}. 

\subsection{Phase-space constraints for vacuum-like emissions}
\label{sec:med_phase_space}

The first effect that we discuss is the reduction of phase space for vacuum-like emissions~\cite{Caucal:2018dla,Mehtar-Tani:2017web}. Sufficiently hard splittings, corresponding to very short formation times,\footnote{We remind the reader that the typical formation time is related to the off-shellness of the emitter and is given by $t_f=2/(z(1-z)p_t\theta^2)$, with $z$ the energy-fraction of the emission and $\theta$ the opening angle of the splitting.} are unmodified by the medium, i.e., their splitting probability is given by Eq.~\eqref{eq:splitting_functions}. Conversely, splittings with longer formation times are sensitive to medium dynamics and their radiation pattern is qualitatively modified due to gluon exchanges with the medium constituents. In particular, assuming that the medium-jet interactions are dominated by multiple soft gluon exchanges, the typical formation time for medium-induced emissions is given by
\begin{equation}
\label{eq:med-formation_time}
t^{\rm med}_{f} = \sqrt{\frac{2\omega}{\hat q}},
\end{equation}
where $\omega$ is the energy of the emission and $\hat q$ is the so-called jet quenching parameter, that depends on the properties of the medium.\footnote{In what follows, we shall implicitly use $\hat q $ as denoting the jet quenching parameter in the adjoint color representation.} Thus, vacuum-like splittings inside the medium ($t_f < L$) must satisfy $t_f>t^{\rm med}_{f}$. These temporal constraints lead to the definition of a so-called veto region for in-medium vacuum splittings~\cite{Caucal:2018dla}
\begin{align}
\label{eq:veto-ps}
   \Theta_{\rm veto} &=\Theta(t_f-t^{\rm med}_{f}) \Theta(L-t_f),
\end{align}
which propagates to the anomalous dimension considered in Eq.~\eqref{eq:gamma_vac}. That is, vacuum evolution is forbidden in $\Theta_{\rm veto}$ and the new anomalous dimensions read (see also~\cite{tracks_paper}):
\begin{align}\label{eq:gamma_med}
   \gamma^{\rm med}(j,\theta) = - \int_0^1 \dd z \, z^{j-1}  \hat P(z)[1- \Theta_{\rm veto}(z,\theta)]\, .
\end{align}
In particular, for the $q\to qg$ channel we find
\begin{align}\label{eq:gamma_med_explicit}
\gamma_{qq}^{\rm med}(j,\theta) &=  \gamma_{qq}^{\rm vac}(j)+ \int_0^1 \dd z \, z^{j-1}  P_{gq}(z)\Theta_{\rm veto}(z,\theta)\, .
\end{align}
\begin{figure}
    \centering
    \includegraphics[width=.49\textwidth]{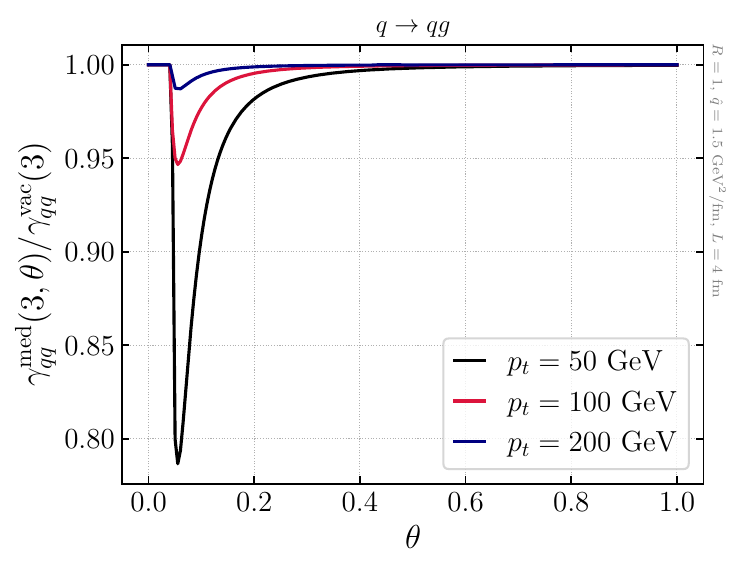}
    \includegraphics[width=.49\textwidth]{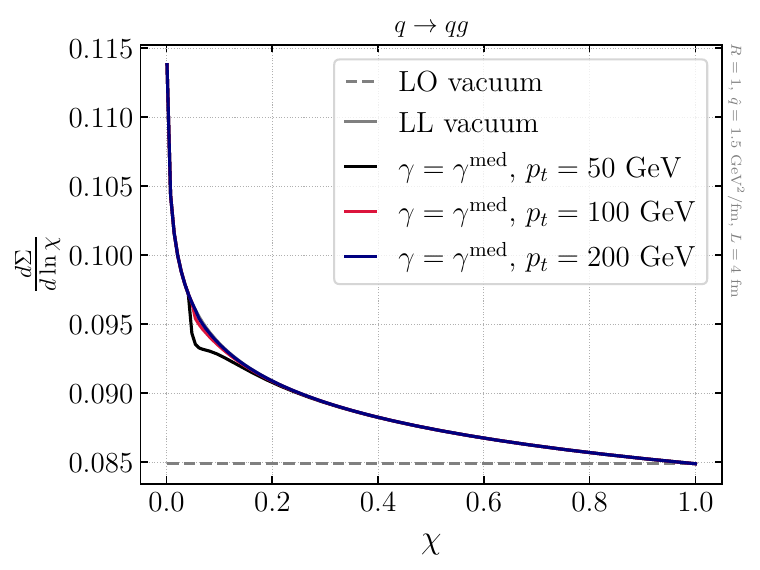}
    \caption{Left: Medium-to-vacuum ratio of the anomalous dimension for $j=3$ as a function of the splitting angle for the $q\to qg$ channel, see Eq.~\eqref{eq:gamma_med_explicit}. Right: the energy-energy correlator in vacuum (both at fixed and all orders) together with the result including the medium-modified anomalous dimensions shown in the left panel.  }
    \label{fig:gammas}
\end{figure}
 An important remark is that the phase-space constraint, $\Theta_{\rm veto}$ is derived in the soft-and-collinear limit. Consequently, the splitting function entering Eq.~\eqref{eq:gamma_med_explicit} corresponds to $P_{gq}(z)=2C_F/z$. In this approximation, the anomalous dimension in vacuum reduces to $\gamma^{\rm vac}_{qq}(j) = -2C_F/(j-1)$. The angular dependence of $\gamma_{qq}(j,\theta)$ is shown in the left panel of Fig.~\ref{fig:gammas} for $j=3$ (the relevant case for the EEC calculation) both in vacuum and medium for different values of the jet $p_t$. The medium parameters are fixed to be $L=4$ fm and $\hat q=1.5$ GeV$^2$/fm, roughly corresponding to central PbPb collisions at LHC energies. 
 We observe that for all values of jet $p_t$ the in-medium anomalous dimension is reduced compared to the vacuum baseline. This can be naturally understood as a result of a reduction of the radiative phase space. In addition, the medium-to-vacuum ratio of anomalous dimensions tends to unity with increasing jet $p_t$, i.e., highly energetic jets undergo a vacuum-like dominated evolution and are less sensitive to the medium scales. A final comment concerns the sharpness on the angular dependence of the in-medium $\gamma_{qq}(3)$, which is due to the step-wise nature of the model describing $\Theta_{\rm{veto}}$. Higher-order perturbative corrections and fluctuations of medium scales are expected to smooth out the boundaries of the resolved phase space. 
 
 To end this section, we show on the right panel of Fig.~\ref{fig:gammas} the impact of the medium-modified anomalous dimensions on the EEC. Note that we have re-scaled the vertical axis by the EEC angle such that the leading order prediction becomes a flat line, as can be deduced from Eq.~\eqref{eq:eec-LO}. Including resummation effects leads to deviations from this flat line, as observed for the solid, gray line in the figure. Also note that in the leading-logarithmic approximation and fixed coupling, the EEC is independent of the jet $p_t$. We find that the EEC slope is mildly modified ($<5\%$ for this choice of parameters) in the moderate angle region after incorporating phase-space constraints on the definition of the anomalous dimensions. 

\subsection{Medium modified splitting function}
\label{sec:med-split-function}
Another source of medium-induced modification to the EEC results from the production of \textit{bremsstrahlung} radiation due to the interactions between hard jet constituents and the medium, leading to  an excess of soft gluons at large angles. In a perturbative approach, this can be captured, at leading order in the strong coupling, by writing the $1\to 2$ cross-section as
\begin{equation}\label{eq:x-section_full}
\frac{\dd^2\cP^{\rm full}}{\dd z \dd \theta} =\frac{\dd^2 \cP^{\rm vac}}{\dd z \dd \theta}+\frac{\dd^2\cP^{\rm med}}{\dd z \dd \theta} \equiv [1+F_\med(z,\theta)] \frac{\dd^2\cP^{\rm vac}}{\dd z \dd\theta}\, ,
\end{equation}
where medium induced radiation contributions are encapsulated inside the $F_{\rm med}$ function and the vacuum term is given by Eq.~\eqref{eq:splitting_fucntions_2}.\footnote{If the strong coupling constant differs in vacuum and medium, the $1\to 2$ cross-section takes the form $\frac{\dd^2\cP^{\rm full}}{\dd z \dd \theta} = [1+\frac{\alpha_s^\med}{\alpha_s}F_\med(z,\theta)] \frac{\dd^2\cP^{\rm vac}}{\dd z \dd\theta}$.} Beyond $\mathcal{O}(\alpha_s)$, this separation between vacuum and medium physics in the splitting function remains to be fully studied, see e.g.~\cite{Arnold:2016kek, Fickinger:2013xwa}. Combining Eq.~\eqref{eq:EEC_def} with Eq.~\eqref{eq:x-section_full} one can write the leading order in-medium EEC as
\begin{align}
	\frac{\dd\Sigma}{\dd\theta\dd p_t} &=  \sum_{\{i,j\}}\int_0^1 \dd z  \left[ z(1-z) \frac{\dd \cP_{ij}^{\rm vac}}{\dd\theta \dd z}	\left(1+ F^{ij}_{\rm med}(\theta,z)\right)\right]   \frac{\dd \sigma_j}{\sigma_j \dd p_t} \, ,
\end{align}
where the last term denotes the jet cross-section that generates the leading parton.\footnote{In other words, $\frac{\dd \sigma_j}{\sigma_j \dd p_t}$ is the hard function. It is omitted in Eq.~\eqref{eq:x-section_full} since, in the vacuum case, it only changes the overall normalization of the EEC distribution. Note that in this work, we are focused on understanding the final state modifications to the EEC, and we do not provide a complete computation of the observable.} In what follows, we will mainly restrict the discussion to $q\to qg$ and $\gamma \to q\bar q$ splittings and postpone the discussion on the role of the initial jet cross-section to the next section. 

Despite substantial progress over the last decades, the medium modification factor $F_{\rm med}$ can only be efficiently computed in particular kinematical limits. The first studies of the in-medium EEC used the so-called `semi-classical' limit~\cite{Dominguez:2019ges} where the outgoing partons are considered to be hard ($z\sim 1/2$). Alternatively, several phenomenological works used the soft limit approximation for the radiated gluon ($z\ll 1$); we shall refer to this as the BDMPS-Z limit~\cite{Zakharov:1996fv,Wiedemann:1999fq,Baier:1994bd,Salgado:2003gb}. As we will discuss below, the BDMPS-Z approach is insufficient for computing the EEC since this observable involves finite energy fractions, e.g. the energy weight in the EEC definition in Eq.~\eqref{eq:EEC_def}. Finally, we will also consider a recent numerical calculation of $F_{\rm med}$ with exact kinematics in the multiple soft scattering approximation~\cite{Isaksen:2023nlr} for the simpler $\gamma\to q \bar q$ channel. So far, this numerical approach is limited to pair production from a photon, which is a subleading process for jet production at the LHC. Nevertheless, this channel captures many of the structures entering other flavor's branchings~\cite{Isaksen:2020npj}, and we will use this exact result to gauge the validity of the above-mentioned approximations, which we follow to discuss in more detail.

\paragraph{Semi-classical approximation:} We consider the outgoing partonic states in a $1\to 2$ process to be very energetic, such that evolution in the medium is mainly dominated by the rotation of their color field, while deflections from the classical trajectory are neglected. In this case, the branching process itself (and not the subsequent evolution in the QGP) mostly controls the final transverse momentum distribution. The expression for the modification factor in this approximation is given by~\cite{Dominguez:2019ges,Isaksen:2020npj} 
\begin{align}\label{eq:Fij}
	F_{\rm med}= \frac{2}{t_f} \left[ \int_0^L \dd t\, \left\{ \int_t^L \frac{\dd t'}{t_f} \,  \cos\left(\frac{t'-t}{t_f}\right) \mathcal{C}_3(t',t) \mathcal{C}_4(L,t')\right\} - \sin\left(\frac{L-t}{t_f}\right) \mathcal{C}_3(L,t) \right]\, ,
\end{align}
where $\mathcal{C}_{3,4}$ denote particular projections of three and four-point correlation functions of in-medium propagators, see~\cite{Dominguez:2019ges,Isaksen:2020npj,Isaksen:2023nlr} for further discussion and details on the calculation of these objects.
For $\gamma \to q\bar q$ and $q\to qg$, the $\mathcal{C}_{3,4}$ correlators can be reduced to relatively simple forms in a large number of colors, $N_c$, limit:
\begin{align}\label{eq:C3_C4}
	\mathcal{C}_{3}(t',t)&=\mathcal{C}_3(t'-t)= e^{-\frac{1}{12} \hat q \theta^2 \zeta(z) (t'-t)^3} \, ,\nn
 \mathcal{C}_4(L,t')&= e^{-\frac{1}{4}\hat q \xi(z) \theta^2 (L-t') (t'-t)^2} + {\rm non-factorizable} \, .
\end{align}
The flavor dependence of the process is encapsulated in the $\zeta$ and $\xi$ functions, which for the two channels under consideration read~\cite{Isaksen:2020npj}~\footnote{We thank Carlota Andrés and Fabio Dominguez for pointing out a typo in Ref.~\cite{Isaksen:2020npj} as due to $z$ and $1-z$ being swapped in the $q\to qg$ case.}: 
\begin{align}
    \xi_{\gamma\to q\bar q} &= z^2+(1-z)^2 \,, \quad  \zeta_{\gamma \to q\bar q} = 1 \, ,\nn 
    \xi_{q\to gq} &= 1-2(1-z)+3(1-z)^2\,, \quad  \zeta_{q \to gq} =     1+(1-z)^2 + \frac{2(1-z)}{N_c^2-1}  \, .
\end{align}
Another point concerning Eq.~\eqref{eq:C3_C4} is the role of the so-called non-factorizable pieces in $\mathcal{C}_{4}$. These terms account for non-trivial color configurations, and in Refs.~\cite{Isaksen:2020npj,Dominguez:2019ges,Blaizot:2012fh} it was argued that their contribution should remain quantitatively small for dense media. In this work, we have explicitly checked that the impact of these non-factorizable corrections in the EEC, for a set of parameters used in the figures shown, is negligible, and, due to their numerical complexity, we neglect them in what follows. Finally, plugging Eq.~\eqref{eq:C3_C4} into Eq.~\eqref{eq:Fij}, the medium modification factor can be compactly written as 
\begin{align} \label{eq:Fmed-hard}
	F_\med= \frac{2}{t_f} \int_0^L \dd t \,\left[4 \frac{1-e^{-\frac{1}{4} \xi \hat q (L-t)\theta^2 t^2 }}{\xi \hat q t_f \theta^2 t^2} \cos\left(\frac{t}{t_f}\right)  - \sin\left(\frac{t}{t_f}\right)  \right] e^{-\frac{1}{12} \hat q \zeta \theta^2 t^3 }\, .
\end{align}%

\begin{figure}
    \centering
    \includegraphics[width=.49\textwidth]{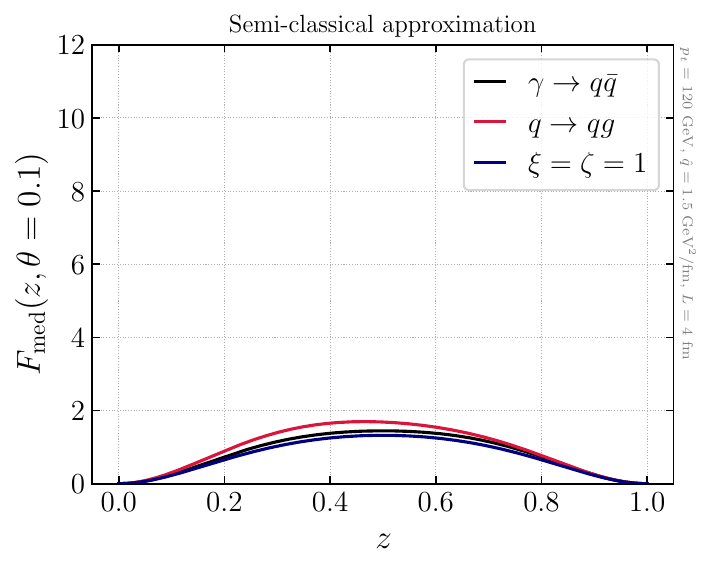}
    \includegraphics[width=.49\textwidth]{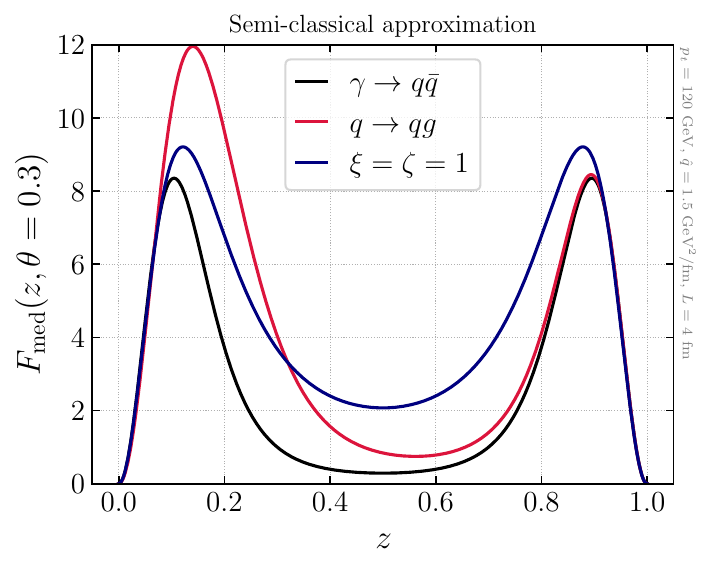}
    \caption{Evaluation of $F_\med$ in the semi-classical approximation following Eq.~\eqref{eq:Fmed-hard} for two different splitting angles, $\theta=0.1$ (left) and $\theta=0.3$ (right), and three different splitting processes.}
    \label{fig:Fmed_z}
\end{figure}

In Fig.~\ref{fig:Fmed_z} we show the evaluation of Eq.~\eqref{eq:Fmed-hard} as a function of $z$ for two values of $\theta$. We consider two physical partonic channels and a toy scenario in which $\zeta = \xi = 1$, which will be used in the Monte Carlo implementation of $F_{\med}$ presented in Sec.~\ref{sec:results-jetmed}. We note a drastic difference in the behavior of $F_{\med}$ between collinear ($\theta=0.1$) and wide-angle splittings ($\theta=0.3$). For small-angle splittings, we observe that the dominant contribution to $F_{\med}$ comes from democratic splittings ($z\sim 1/2$). In contrast, the wide-angle limit of $F_{\med}$ corresponds to asymmetric splittings in which either the daughter parton ($z\to 0$) or the emitter ($z\to 1$) are soft. This is an important point since very asymmetrical configurations can not be described using the semi-classical approximation. Consequently, as the angle increases, one expects the values obtained for $F_{\med}$ to be increasingly less accurate. We anticipate that, counter-intuitively, these small $z$ branchings are precisely the ones that dominate the wide-angle part of the jet EEC at the leading order. We observe this to be always the case when using the semi-classical all approximation, even after including the energy suppression factor entering the EEC definition. This calls for a more accurate calculation or better model for the soft sector. 

Beyond the semi-classical approximation, the $F_{\rm med}$ factor receives corrections related to \textit{quantum diffusion} in the transverse plane, which can be captured in a power series expansion in $k_\perp^2 \Delta t/p_t$~\cite{Altinoluk:2014oxa}, with $k_\perp$ ($\Delta t$) a typical transverse momentum  (time) scale. When $k_\perp^2$ becomes commensurate with the typical momentum transfer from the medium $\hat q L$, and $\Delta t\sim L$, one can estimate that the semi-classical approximation fails if $\min(z,1-z) p_t \ll \hat q L^2/2\equiv \omega_c$, assuming $p_t\gg \omega_c$.

\paragraph{Soft approximation:} We now explore the kinematical limit where $z\to 0$ or $z\to 1$. This regime corresponds to the emission of very soft radiation along with a hard, nearly eikonal parton. We focus on the $q\to qg$ splitting process. The $F_{\med}$ modification factor can be calculated under these conditions in the BDMPS-Z limit for $z\to 0$, or in the opposite limit when $z\to 1$~\cite{Apolinario:2012vy}. For the simple medium model considered above, the in-medium emission probability reads (see e.g.~\cite{Barata:2021wuf} for details): 
\begin{align}
\label{eq:pmed-bdmps}
\frac{\dd^2\cP^{\rm med}_{z\to 0}}{\dd z \dd \theta} = 2\pi  \, \theta p_t \omega^2\, \frac{\dd^2\mathcal{P}^{\rm med}_{z\to 0}}{\dd\omega \dd^2\k} \;,
\end{align}
where we have implicitly used that the spectrum does not depend on the azimuthal angle, with
\begin{align}\label{eq:bdmps}
\frac{\dd^2\mathcal{P}^{\rm med}_{z\to 0}}{\dd\omega \dd^2\k}= \frac{8\alpha^{\med}_s C_F}{4\pi^2  \omega} {\rm Re} \Bigg[ & \int_{0}^L \dd t~\Omega\cot(\Omega t) \, \frac{e^{-\frac{k_\perp^2}{\hat q (L-t)-2 i \omega\Omega\cot(\Omega t)}}}{\hat q (L-t) - 2 i\omega \Omega\cot(\Omega t)} \nn 
&- \frac{1}{k_\perp^2}\Big(1-e^{-\frac{i k_\perp^2}{2\omega\Omega \cot(\Omega L)}}\Big) \Bigg] \, ,
\end{align}
where $\Omega=\frac{(1-i)}{2} \sqrt{\hat q/\omega}$, $|\k|=k_\perp$, and $\omega=z p_t$. Note that the strong coupling in the medium $\alpha_s^{\rm med}$ is evaluated at $(\hat q \omega)^{1/4}$, which is the typical transverse momentum scale for medium-induced emissions \cite{Blaizot:2012fh}. The case $z\to 1$ can be obtained from Eq.~\eqref{eq:bdmps} by crossing symmetry and reads~\cite{Apolinario:2012vy} 
\begin{align}\label{eq:bdmps-hard}
	\dd^2\cP^{\rm med}_{z \to 1} = \frac{1}{2} 	\left.\dd^2\mathcal{P}^{\rm med}_{z\to 0}\right\vert_{z\to 1-z ,  \hat q \to   \hat q_{\rm F} }\,,
\end{align}
where the correct jet quenching parameter is now in the fundamental representation, i.e., $C_A\hat q_{\rm F}=C_F \hat q $. The overall $1/2$ factor accounts for the transformation of $P_{gq}$ between the hard and soft limits, see Eq.~\eqref{eq:splitting_functions}.

\begin{figure}
    \centering
    \includegraphics[width=.49\textwidth]{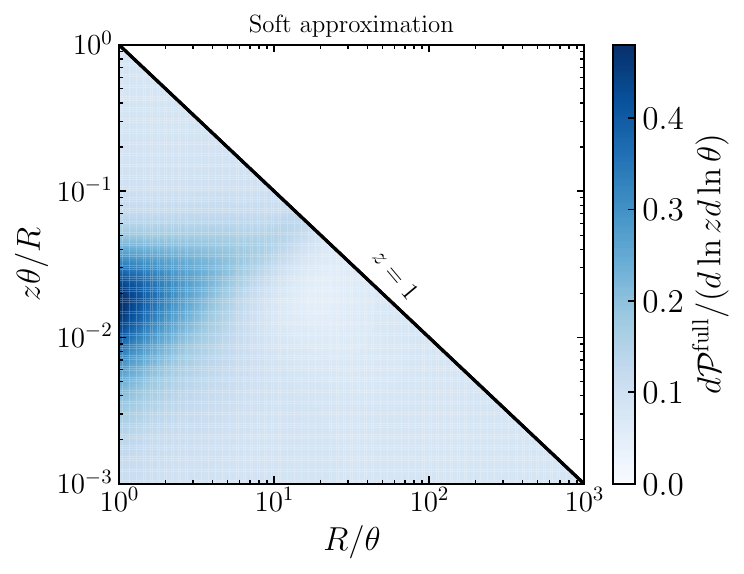}
    \includegraphics[width=.49\textwidth]{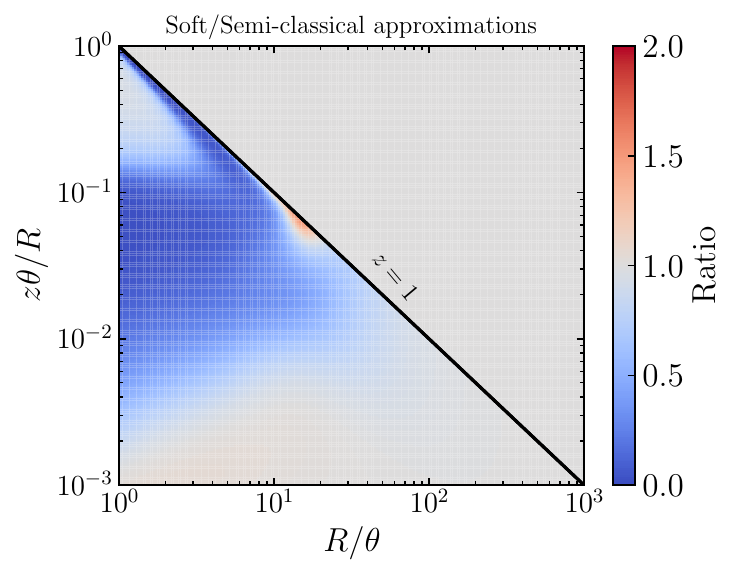}
    \caption{Lund-plane representation of the differential probability distribution given in Eq.~\eqref{eq:x-section_full}. Left: soft approximation for the medium splitting probability as given by Eq.~\eqref{eq:bdmps}. Right: ratio between the soft (Eq.~\eqref{eq:bdmps}) and semi-classical approximations (Eq.~\eqref{eq:Fmed-hard}) to the medium kernel. In both plots, we choose: $\hat q=1.5$ GeV$^2$/fm, $L=4$ fm, $p_t=120$ GeV, $R=1$, $\alpha_s=0.1$ and $\alpha^{\rm med}_s=0.24$. }
    \label{fig:Fmed_lund}
\end{figure}

We show in the left panel of Fig.~\ref{fig:Fmed_lund} the splitting probability density defined in Eq.~\eqref{eq:x-section_full} using the soft approximation for the medium kernel (Eq.~\eqref{eq:pmed-bdmps}). Note that in this Lund-plane representation, the vacuum splitting probability reduces to a constant ($2\alpha_s C_F/\pi$) at double-logarithmic accuracy. Therefore, any enhancement or depletion in this graph can be attributed to medium effects. In particular, the enhancement observed for soft, wide angle splittings corresponds to the characteristic scale of the medium $z\theta p_t=\hat q L $. The right panel of Fig.~\ref{fig:Fmed_lund} shows the ratio between this soft approximation and the semi-classical limit discussed above. We would like to remark that the quantitative interpretation of this ratio is delicate. Nevertheless, we observe clear differences between these two approximations of the in-medium matrix element in almost all regions of the radiative phase-space. In particular, we confirm that the semi-classical approximation does not capture the BDMPS-Z result neither when $z\to 0$ nor when $z\to 1$. Unfortunately, none of these approximations can be compared to the exact matrix element since the numerical method proposed in Ref.~\cite{Isaksen:2020npj} for $\gamma\to q\bar q$ is not yet available for $q\to qg$. 

\begin{figure}
    \centering
    \includegraphics[width=.49\textwidth]{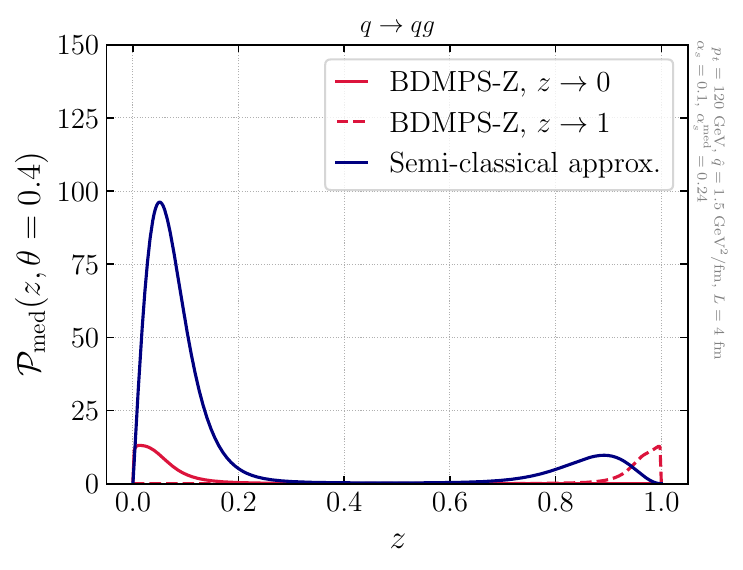}
    \includegraphics[width=.49\textwidth]{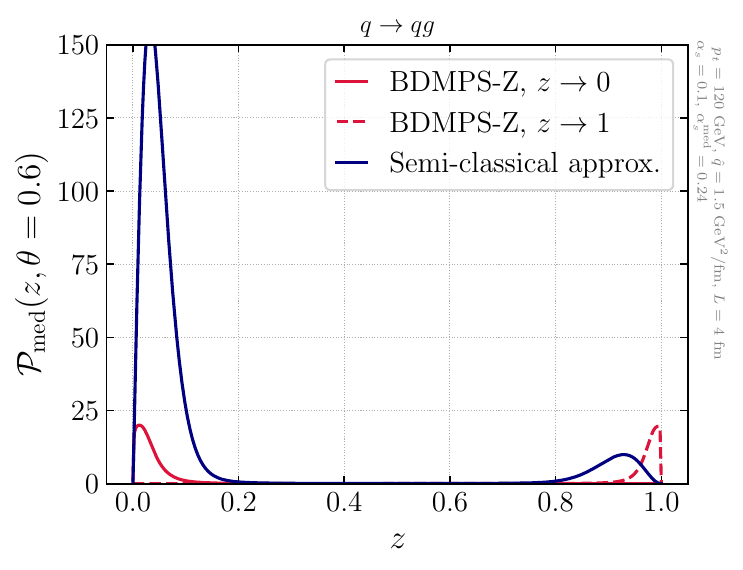}
    \caption{Medium splitting probability in $q\to qg$ as a function of $z$ for two different splitting angles: $\theta=0.4$ (left) and $\theta=0.6$ (right). We show two different approximations for the splitting kernel: soft (or BDMPS-Z) both in the $z\to 0$ and $z\to 1$ limits, and semi-classical. }
    \label{fig:Pmed_bdmps-vs-fmed}
\end{figure}

To further analyze the differences between the semi-classical and soft limits, we plot the in-medium scattering probability for fixed-angles in Fig.~\ref{fig:Pmed_bdmps-vs-fmed}. Note that these plots contain the full splitting probability (vacuum + medium) and not just the medium part, as was done in Fig.~\ref{fig:Fmed_z}. Both panels focus on the large-angle regime, the region of interest for medium-modifications to the energy-energy correlator. In the BDMPS-Z case, we show both the soft ($z\to 0$) and hard gluon ($z\to 1$) limits given by Eqs.~\eqref{eq:bdmps} and \eqref{eq:bdmps-hard}, respectively. We find that the large-angle regime is fully dominated by asymmetric splittings whose description clearly differs when taking the soft or the semi-classical limit. The imprint of this mismatch in the EEC itself will be discussed in Sec.~\ref{sec:results_analytics}.    

\paragraph{Heuristic interpolation scheme:} After separately studying three approximations of the exact in-medium matrix element in different kinematic regimes ($z\sim 1/2$, $z\to 0$ and $z\to 1$), here we propose an \textit{ansatz} for the cross-section that simultaneously covers these three kinematical limits. Assuming $p_t \gg \omega_c$, we consider the following interpolating scheme: 
\begin{align}
\label{eq:ansatz}
\dd^2\cP^{\rm med} &= \dd^2\cP^{\rm Eq.~\eqref{eq:Fmed-hard}}\,  \Theta(\min(z,1-z)p_t -\omega_c) \nn 
&+   \dd^2\cP^{\rm Eq.~\eqref{eq:bdmps}}\, \Theta(\omega_c-zp_t ) + \dd^2\cP^{\rm Eq.~\eqref{eq:bdmps-hard}}\, 	\Theta( \omega_c-(1-z)p_t)\, ,
\end{align}
which splits the phase-space in terms of the emission's energy. That is, the first line uses the semi-classical approximation in a region of symmetric energy sharing, while the second line describes two regions where most of the energy is carried away by one of the daughter particles. Note that this form is introduced to gauge how much the most asymmetrical branching region affects the in-medium EEC. This is a relevant question since previous studies of the EEC used the semi-classical approximation in the full $z$-range, even if it does not capture the correct $z\to 0$ and $z\to 1$ limits, as shown in Figs.~\ref{fig:Fmed_lund} and \ref{fig:Pmed_bdmps-vs-fmed}. As we will show next, these asymmetric configurations  actually give a sizable contribution to the EEC at large angles in this particular model.
Finally, we would like to remark that the \textit{ansatz} introduced in Eq.~\eqref{eq:ansatz} is not meant as a rigorous, or even improvable, way to obtain a more accurate in-medium cross-section. 

\subsection{Energy loss model for two parton system}
\label{sec:e-loss}

Finally, we consider the effect of energy loss on the in-medium cross-section. From a phenomenological point of view, energy loss is expected to be the main source of medium modifications to jet observables. On the theory side, a complete description of the energy loss mechanism to all-orders remains challenging (see Refs.~\cite{Mehtar-Tani:2017ypq,Mehtar-Tani:2017web} and references therein for recent progress). The treatment of energy loss presented here closely follows that introduced in Ref.~\cite{Mehtar-Tani:2017web} and used in other studies both about the EEC~\cite{Barata:2023vnl} and beyond~\cite{Takacs:2021bpv,Caucal:2021cfb,Mehtar-Tani:2021fud,Pablos:2022mrx}.  

More concretely, we consider a model for jet energy loss based on the quenching weight approximation~\cite{Baier:2001yt,Salgado:2003gb}. Without loss of generality, let us consider a $q$-initiated jet. The first step is to approximate the in-medium cross-section, $\dd\sigma_q$,  as a $p_t$-shifted version of the vacuum one, $\dd\sigma_q^{\rm vac}$, i.e.,
\begin{align}~\label{eq:master_energy_loss}
	\frac{\dd\sigma_q^{\rm med}}{\dd p_t \dd\theta} &=  \int_0^\infty \dd\varepsilon \, D_q(\varepsilon)  \frac{\dd\sigma_q^{\rm vac}}{\dd p'_t \dd\theta}\Bigg\vert_{p'_t=p_t+\varepsilon}  \, ,
\end{align}
where $D_q(\varepsilon)$ is a probability distribution describing the transfer of $\varepsilon\ll p_t$ energy from the quark to the QGP by means of out-of-the cone emissions. Next, the quenching weight approximation exploits the steeply falling nature of the vacuum cross-section ($\dd\sigma_q\sim \dd p_t^2/p_t^n $ with $n\approx 6$) to approximate Eq.~\eqref{eq:master_energy_loss} as 
\begin{align}~\label{eq:quenching_weight_energy_loss}
	\frac{\dd\sigma_q^{\rm med}}{\dd p_t \dd\theta} &\approx \frac{\dd\sigma_q^{\rm vac}}{\dd p_t \dd \theta}  \int_0^\infty \dd \varepsilon \, D_q(\varepsilon) e^{-\frac{n \varepsilon}{p_t}}  \equiv Q_q(p_t) \frac{\dd\sigma_q^{\rm vac}}{\dd p_t \dd \theta}  \, ,
\end{align}
where $Q_q(p_t)$ is usually referred to as the (quark) quenching weight and represents the Laplace transform of the single parton energy loss probability.

As we have already mentioned, the single parton energy loss probability distribution is, in general, a highly complex object (including non-perturbative ingredients) and its full description is still not understood. Focusing on a partonic picture, a closed form for $Q_q(p_t)$ can be achieved by assuming that multiple gluon emissions are independent, a reasonable assumption when the jet energy is transported at large angles via soft gluons~\cite{Blaizot:2013hx,Baier:2000sb,Berges:2020fwq}. Within this approximation, one can write the single body quenching weight as~\cite{Baier:2001yt}
\begin{align}\label{eq:Qi_1}
 Q_i = \exp \left[-\int \dd \omega \int \dd^2\k\, \frac{\dd\cP^{\rm med}_i}{\dd\omega \dd^2\k} \left(1- e^{-\frac{n \omega}{p_t}}\right) \right]\, ,
\end{align}
and we use the BDMPS-Z form of the in-medium cross-section (since it captures the production of soft radiation) as given by Eq.~\eqref{eq:bdmps}. At this point, it is important to remark that there are two distinct physical regimes in the medium-induced cascade that are responsible for energy loss. To facilitate the discussion, we split Eq.~\eqref{eq:Qi_1} as:
\begin{align}
\label{eq:Qi_2}
 Q_i &= \exp \left\{-	\int_T^{\omega_s} \dd\omega \int \dd^2\k \frac{\dd\cP^{\rm med}_i}{\dd\omega\dd^2\k} \left(1- e^{-\frac{n\omega}{p_t}} \right)	 +  \int_{\omega_s}^\infty \dd\omega \int \dd^2\k\frac{\dd\cP^{\rm med}_i}{\dd\omega \dd^2\k} \left(1- e^{-\frac{n\omega}{p_t}}\right)	\right\}\nn 
 &\equiv Q_i^{\rm mini-jets} \times Q_i^{\rm pert.} \,, 
\end{align}
and introduce the characteristic scale $\omega_s\equiv \left(\frac{\alpha^{\rm med}_s N_c}{\pi}\right)^2 \omega_c$~\cite{Blaizot:2013hx}. Physically, $\omega_s$ corresponds to the energy scale for which the in-medium emission probability becomes order one, and the medium-induced cascade develops a turbulent behavior. Soft gluon emissions with $T<\omega<\omega_s$ thermalize quickly, and their perturbative description breaks down. Further, their emission rate becomes independent of transverse momentum broadening since they occur at large angles. In this mini-jet dominated regime, we can therefore write the quenching weight as:
\begin{align}
Q_i^{\rm mini-jets} &= \exp \left[-	\int_T^{\omega_s} \dd\omega  \frac{\dd\cP^{\rm med}_i}{\dd\omega} \left(1- e^{-\frac{n\omega}{p_t}} \right)\right],~\text{with}  ~\frac{\dd\cP^{\rm med}_i}{\dd\omega} = \frac{2\alpha^{\med}_s C_i}{\pi} 	\sqrt{\frac{\omega_c}{2\omega^3}} \;,
\end{align}
where we have used the $\omega\ll\omega_c$ limit of the medium-induced energy spectrum. After integration, we find 
\begin{align}
	 Q_i^{\rm mini-jets} &=  \exp \Bigg\{-\frac{2\alpha^{\med}_s C_i}{\pi}  \Bigg[ \sqrt{\frac{2\omega_c}{T}} \left(1-e^{-\frac{nT}{p_t}}\right) -  \sqrt{\frac{2\omega_c}{\omega_s}} \left(1-e^{-\frac{n \omega_s}{p_t}}\right) \nn 
	&+ \sqrt{\frac{2 \pi \omega_c n}{p_t}} \left(\erf\left(\sqrt{\frac{\omega_s n}{p_t}}\right) -\erf\left(\sqrt{\frac{nT}{p_t}}\right)  \right)\Bigg] \Bigg\}\, .
\end{align}
The second contribution to $Q_i$ in Eq.~\eqref{eq:Qi_2}, $Q^{\rm pert.}_i$, captures the transport of energy out of the jet cone due to the emission of semi-hard (perturbative) gluons with $\omega\gg \omega_s $ that do not instantly thermalize as was the case for the mini-jets. Note that we describe the in-medium emission of these gluons with the BDMPS-Z spectrum in all the frequency ranges, i.e., even when $\omega>\omega_c$. Strictly speaking, an accurate description of gluons with $\omega>\omega_c$ requires going beyond the multiple soft scattering approximation for the in-medium elastic scattering rate and accounting for its Coulomb-like tail ~\cite{Schlichting:2021idr,Mehtar-Tani:2019tvy,Feal:2018sml,Caron-Huot:2010qjx}. Nevertheless, we keep the BDMPS-Z approximation, which yields a $1/\omega^3$ suppression for $\omega>\omega_c$  to facilitate analytic calculations. Further, when $\omega_s\ll \omega\ll \omega_c$, the double differential soft gluon spectrum reduces to \cite{Mehtar-Tani:2016aco,Caucal:2021cfb}
\begin{align}
\label{eq:bdmps-factorize}
	(2\pi)^2 \frac{\dd\cP^{\rm med}_i}{\dd\omega \dd^2\k}
 \approx  \frac{\alpha^{\med}_s C_i}{\pi} \sqrt{\frac{\hat q }{\omega^3}} \frac{4\pi}{\hat q} \Gamma_0\left(\frac{\k^2}{\hat q L}\right)\, .
\end{align}
Plugging Eq.~\eqref{eq:bdmps-factorize} into \eqref{eq:Qi_2} and imposing that all gluons are outside the cone, i.e., $k>2\omega R$~\footnote{To avoid boundary effects, we consider out-of-the cone emissions to satisfy $\theta>2 R$.}
\begin{align}
	Q_i^{\rm pert.}	&=  \exp \Bigg\{- \frac{\alpha^{\med}_s C_F}{\pi} \sqrt{\frac{2\omega_cn} {p_t}} \,  I_\alpha\left(\frac{n \omega_s}{p_t}\right)\Bigg\}\, ,
\end{align}
where $\alpha \equiv  \left(\frac{2p_t R}{\sqrt{\hat q L}n}\right)^2$ and
\begin{align}
I_\alpha \left(\frac{n \,  \omega_s}{p_t} \right) =  \int_{\frac{n \omega_s}{p_t}}^\infty \dd x \,\frac{1- e^{-x}}{\sqrt{x^3}}\left(e^{-x^2 \alpha}-\alpha x^2 \Gamma_0(\alpha x^2)\right)\,. 
\end{align}

An important remark is that in the large radius limit, $\alpha\to \infty$ and thus $Q_i^{\rm pert.}\to 1$. This case corresponds to the limiting scenario considered in e.g.~\cite{Andres:2023xwr}, where it is assumed that energy loss effects are absent since all radiation is recovered inside the jet cone. Nonetheless, we note that even in this ideal asymptotic limit, there are energy loss contributions analogous to the ones captured by $Q^{\rm mini-jets}$, which can not be neglected \textit{a priori}.

So far, we have described the quenching weight energy loss prescription for a single color charge propagating in the medium. Since we want to compute the EEC (at least) at LO accuracy, we need the two body energy quenching weight. Such an object has a much richer structure due to the complex color pattern present in multi-gluon processes, for a more detailed discussion, see e.g.~\cite{Mehtar-Tani:2017ypq}. In this work, we employ a simple model for the two-body energy loss, described by the interpolating form 
\begin{align}\label{eq:eloss}
	Q_{ij}(p_t,\theta,z) &=  Q_i(p_t,R) (1-\Theta_{\rm res}) + Q_i(p_t,R) Q_j(p_t,R) \Theta_{\rm res}\, ,
\end{align}
where $\Theta_{\rm res}$ denotes the phase space where the $1\to 2$ branching is resolved by the medium and it is given by 
\begin{equation}
\label{eq:res-ps}
\Theta_{\rm res} = \Theta(\theta-\theta_c)\Theta(t^\med_f-t_f), \quad {\text {with}} \quad \theta_c = \frac{2}{\sqrt{\hat q L^3}}\, .
\end{equation}
Notice that in this notation, $i$ corresponds to the flavor of the parent particle. Also, since we are working at LO accuracy, we neglect radiative corrections to the bare quenching weights introduced before~\cite{Mehtar-Tani:2017web}. 
The first factor in Eq.~\eqref{eq:eloss} ensures that only sufficiently wide angle emissions, where the outgoing states are resolved as separate color charges, can lose energy independently~\cite{Dominguez:2019ges, Mehtar-Tani:2017ypq,Barata:2021byj,Casalderrey-Solana:2011ule,Mehtar-Tani:2011lic,Mehtar-Tani:2012mfa,Casalderrey-Solana:2012evi}; the second piece ensures that long-lived gluon fluctuations are not included because they never decohere from the parent parton.

\subsection{Summary of the results}

Having individually described the different modifications that affect EEC computation in-medium, we finalize the discussion by piecing together all the different elements. Neglecting the initial jet cross-section, the LL medium modified EEC for the $q\to qg$ channel reads
\begin{align}\label{eq:final_qqg}
\frac{\dd\Sigma^{q\to qg}}{\dd \chi} &=  \frac{\bar\alpha}{\chi}\int_0^1 \dd z\, z(1-z)  P_{gq}(z) \left(  \frac{(1-\Theta_{\rm veto})^{gq}}{\chi^{-\bar \alpha \gamma_{qq}^{\rm med}(3,\chi)}} + F^{gq}_{\rm med}(\chi,z) \right) Q_{qg}(p_t,\chi,z) \nn 
&+ \frac{\bar\alpha}{\chi}\int_0^1 \dd z\, z(1-z)  P_{qq}(z) \left(  \frac{(1-\Theta_{\rm veto})^{qq}}{\chi^{-\bar \alpha \gamma_{qq}^{\rm med}(3,\chi)}} + F^{qq}_{\rm med}(\chi,z) \right) Q_{gq}(p_t,\chi,z) \nn 
&\approx \frac{2\bar\alpha}{\chi}\int_0^1 \dd z\, z(1-z)  P_{gq}(z) \left(  \frac{(1-\Theta_{\rm veto})^{gq}}{\chi^{-\bar \alpha \gamma_{qq}^{\rm med}(3,\chi)}} + F^{gq}_{\rm med}(\chi,z) \right) Q_{qg}(p_t,\chi,z)\, ,
\end{align}
where in the last line we have used that, up to the definition of the energy fraction, the two partonic channels are the same. We note that in our description, they can differ, for example, in the definition of the color representation of the jet quenching parameter. In what follows, we shall ignore these distinctions, which have a small numerical impact. We also set $\Theta^{gq}_{\rm{veto}}\approx 0$ in the fixed order term, since the phase-space constraint is derived in the soft limit. Nonetheless, we conducted a numerical check and found that including the constraint derived in the soft radiation limit at leading order accuracy, with a naive extension to finite kinematics, does not significantly change the observable in a qualitative way.

The equivalent expression for $\gamma \to q\bar q$ reads 
\begin{align}\label{eq:final_gammaqq}
	\frac{\dd\Sigma^{\gamma\to q\bar q}}{ d\chi} &=  \frac{\bar \alpha}{\chi } \int_0^1 \dd z \,  z(1-z)   P^\gamma_{q\bar q}(z)\left(1+ F^{q\bar q}_{\rm med}(\chi,z)\right)  Q_{\gamma \to \bar q q}(p_t,\chi,z)\, ,
\end{align}
where, for consistency, we disregard LL resummation for this channel since it will involve off-diagonal terms in the (QED) anomalous dimension matrix. We have also adapted the notation for the quenching weight to make clear that the parent parton is a photon. The $\gamma\to q\bar q$ vacuum splitting function reads for $n_f$ light quark flavors\footnote{In what follows we use $n_f=1$.}
\begin{align}
P^\gamma_{ q \bar q} =  n_f  (z^2+(1-z)^2)	\, .
\end{align}


\section{In-medium results for the EEC}\label{sec:numerics_EEC}
In this section, we provide a quantitative study of the jet EEC, following the derivations presented in the previous sections. We first discuss the results obtained using the semi-analytic formulas for the $q\to qg$ channel, i.e., Eq.~\eqref{eq:final_qqg}. We then compute the exact $\mathcal{O}(\alpha_s)$ result for $\gamma\to q\bar  q$ using the publicly available numerical routines introduced in Ref.~\cite{Isaksen:2023nlr}. The exact numerical result is compared against different semi-analytic estimates discussed in Sec.~\ref{sec:med-split-function}. Finally, we present a Monte Carlo study, using the {\tt JetMed} MC, modified to account for balanced (i.e., $z\sim 1/2$) branchings in the medium.

\subsection{Semi-analytic results for $q\to q g$ splittings}
\label{sec:results_analytics}

\begin{figure}
    \centering
    \includegraphics[width=.49\textwidth]{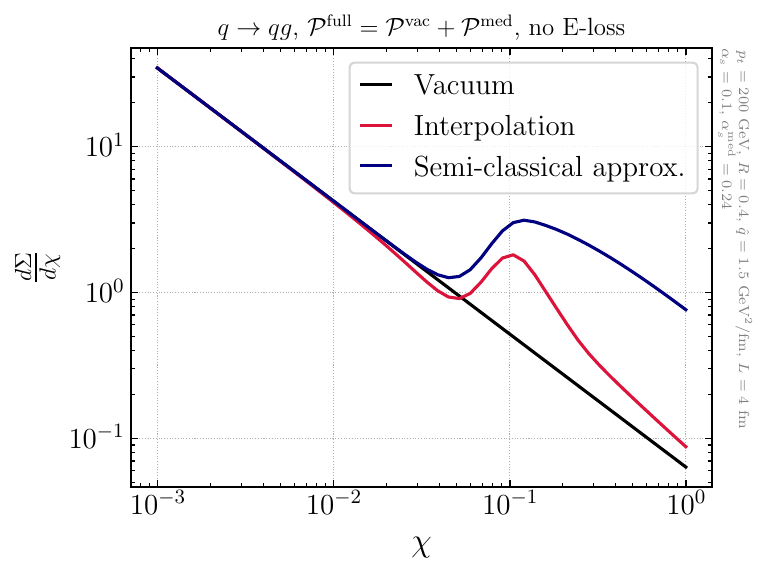}
    \includegraphics[width=.49\textwidth]{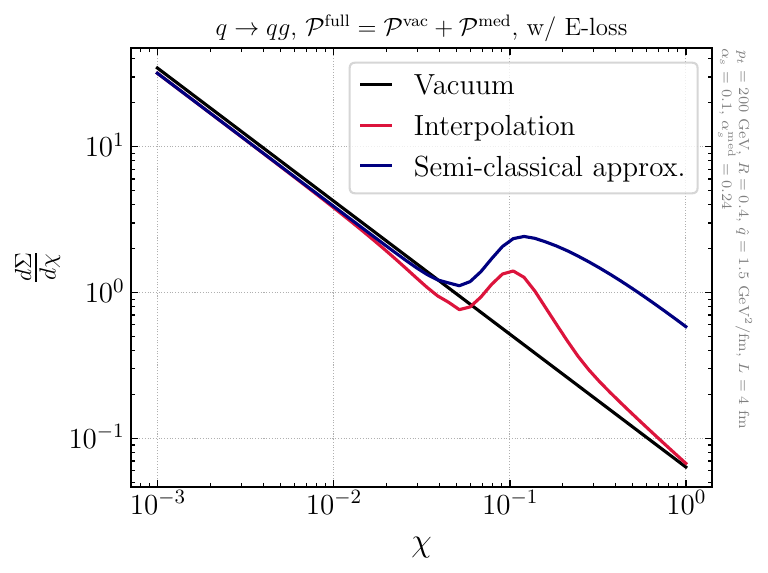}
    \caption{Jet EEC for the semi-analytic models discussed in the main text following Eq.~\eqref{eq:final_qqg}. The left panel neglects energy loss effects, while the right one includes the two-prong energy loss model introduced in the main text.}
    \label{fig:semi_analytics}
\end{figure}

In Fig.~\ref{fig:semi_analytics}, we show the results for the evaluation of Eq.~\eqref{eq:final_qqg} for three different scenarios: pure vacuum (black), in-medium using the semi-classical approximation (Eq.~\eqref{eq:Fmed-hard}) (blue) and the result for the interpolation \textit{ansatz} introduced in Eq.~\eqref{eq:ansatz} (red). We show the results for the case where energy loss is neglected, i.e., $Q_q=Q_g=1$, on the left-hand side, while the right-hand side plot includes the full two parton quenching weight introduced in Eq.~\eqref{eq:eloss}. We consider jets with a radius $R=0.4$ and $p_t=200$ GeV, while the medium parameters follow the choice made in Fig.~\ref{fig:Fmed_z}, with $T=0.3$ GeV. Note that for this set-up $\omega_c=60$ GeV. 

The vacuum result is a straight line with a slope controlled by the anomalous dimension. Regarding the medium curves, let us first focus on the case where energy loss is neglected. For small angles, all curves overlap, meaning that the EEC is dominated by vacuum evolution. Medium effects appear at large angles, in accordance with previous studies~\cite{Andres:2022ovj,Andres:2023xwr}. However, the behavior of the two models for the in-medium matrix element under consideration is qualitatively distinct. When only semi-hard in-medium splittings are considered, we observe a significant enhancement of the EEC, resulting from extra \textit{bremsstrahlung} radiation at wide angles. The curve is characterized by a plateau at the widest angles, while there is a sharp rise at a characteristic scale ($\chi\simeq 0.02$), related to the coherence angle $\theta_c$~(see Eq.~\eqref{eq:res-ps})~\cite{Andres:2022ovj}. In contrast, the interpolation formula first leads to a suppression of the EEC, followed by an enhancement around the same scale as in the semi-classical result. The suppression is driven by the BDMPS-Z spectrum (which can become negative), while the enhancement is again generated by the hard splitting contribution. More importantly, we find that the EEC for $\chi>0.1$ is almost an order of magnitude smaller than the semi-classical result. That is, despite the energy weighting in the observable definition, the regime around the jet boundary is strongly affected by the modeling of very imbalanced splittings ($z\to 0$ or $1$). 

A natural question at this point is whether the previous conclusion might be an artifact due to the over-simplistic interpolation formula that we have chosen. To demonstrate that this is not the case, we have performed three independent checks that do not include the BDMPS-Z contribution. On the one hand, we restricted the range of frequencies over which the EEC is integrated to those for which the semi-classical approximation is valid, i.e., $\omega>\omega_c$. In this case, the EEC falls exactly on top of the vacuum curve at large angles. Alternatively, we have also calculated a groomed version of the observable in which we imposed that $z>z_{\rm cut}=0.1$. Again, the enhancement at large angles disappeared in this scenario. Finally, we computed the EEC using higher-power energy suppression factors, i.e., taking $n>1$ in Eq.~\eqref{eq:eec-definition}.\footnote{Note that the observable becomes collinear unsafe when $n>2$. We discuss an alternative definition of the EEC that allows for any energy weight in Sec.~\ref{sec:LEEC}} The results obtained using the semi-classical approximation showed that raising $n$ leads to a qualitative modification in the shape of the distribution on top of the expected overall suppression. This again indicates a clear dependence on soft branchings. These results might seem counter-intuitive since this class of observables is designed to be insensitive to such soft gluon radiation. However, they indicate that currently available approximations to the in-medium matrix element are insufficient to capture the correct behavior of the in-medium EEC throughout the full kinematic range. Unfortunately, this prevents any qualitative or quantitative interpretation of upcoming data based on these leading-order results. 

Turning to the right panel of Fig.~\ref{fig:semi_analytics}, we observe that the inclusion of energy loss leads to an overall suppression of the EEC. When both outgoing partons are resolved as independent charges, i.e., at angles larger than $\theta_c\simeq 0.04$, this suppression is more prominent, and thus energy loss competes with the modification to the splitting function. The exact balance between these two effects is highly model-dependent, and we can not make a quantitative statement regarding which effect dominates. Nonetheless, we observe that once very soft radiation is removed from the semi-classical approximation, energy loss seems to dominate the observable. In particular, using the EEC to determine the transition between coherent and decoherent jet evolution in the medium is far more involved than the naive $\mathcal{O}(\alpha_s)$ calculation might suggest. We note that the sharp transition at $\chi\sim 0.3$ is due to the step function form used in $\Theta_{\rm res}$; in a more realistic model, this would be smeared due to varying medium parameters and a more accurate treatment of the phase space, see e.g.~\cite{Barata:2023vnl} for a different energy loss prescription for the EEC.

\subsection{Leading order results for $\gamma \to q \bar q$ splittings}
\label{sub:LO-exactFmed}
\begin{figure}
    \centering
    \begin{subfigure}{0.49\textwidth}
        \includegraphics[width=\linewidth]{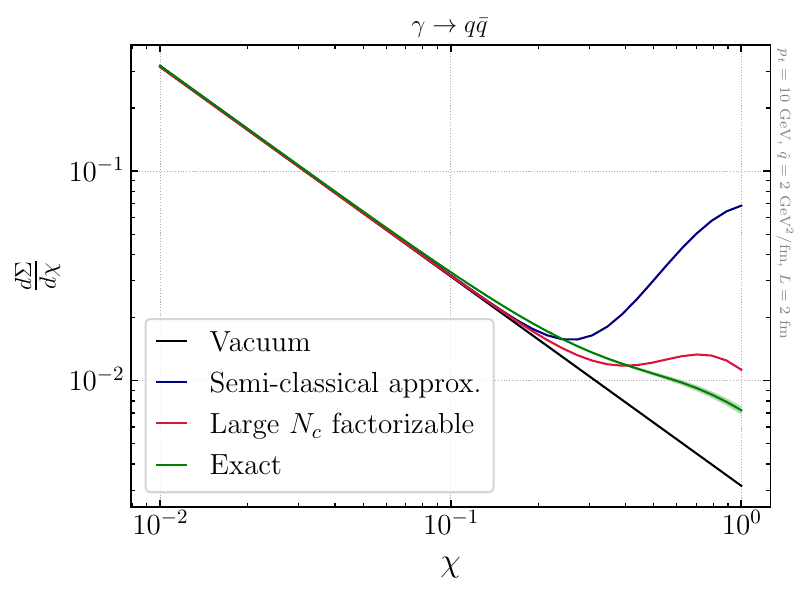}
    \end{subfigure}
    \begin{subfigure}{0.49\textwidth}
        \includegraphics[width=\linewidth]{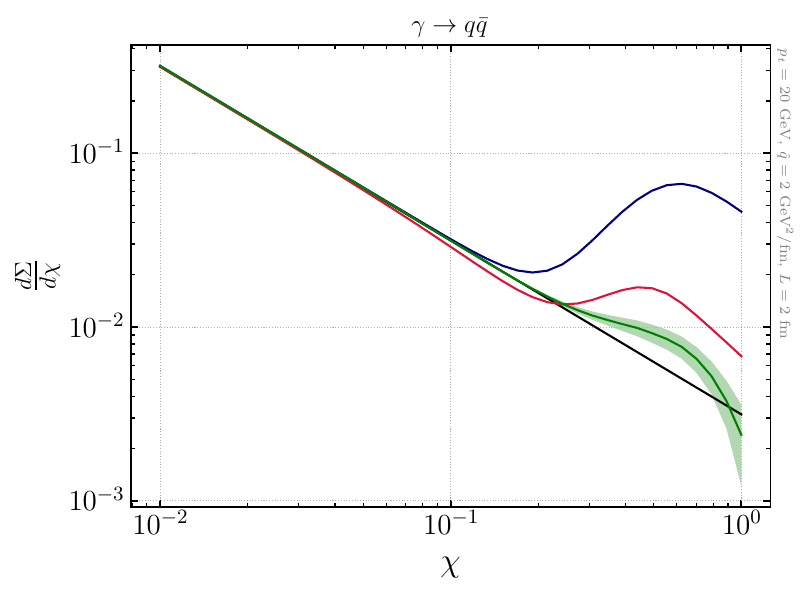}
    \end{subfigure}

    \begin{subfigure}{0.49\textwidth}
        \includegraphics[width=\linewidth]{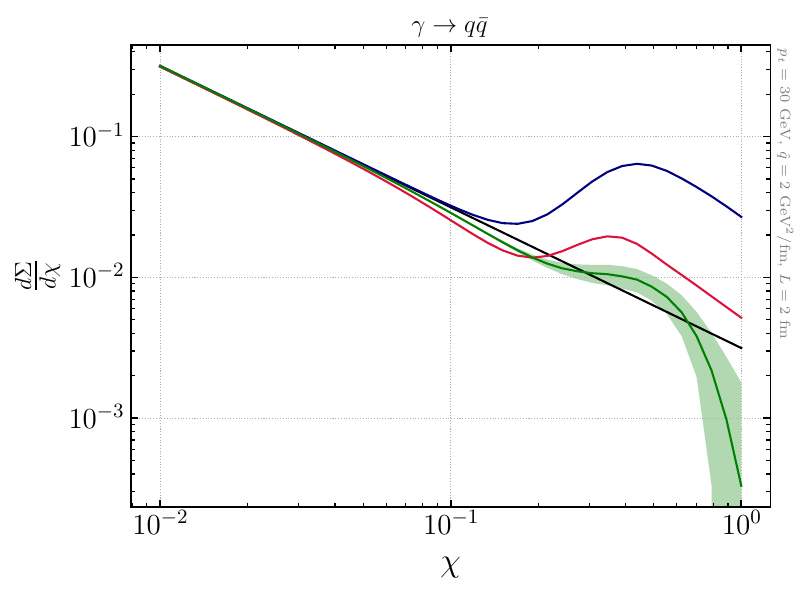}
    \end{subfigure}
    \begin{subfigure}{0.49\textwidth}
        \includegraphics[width=\linewidth]{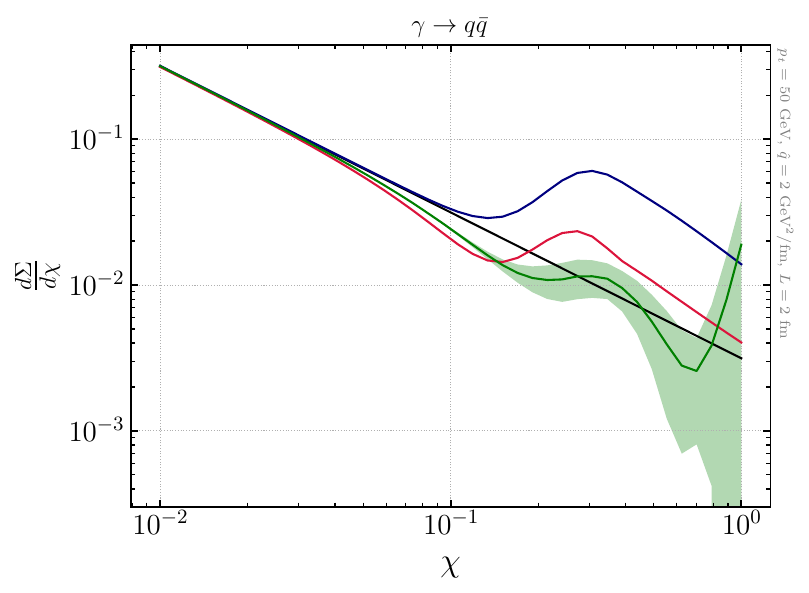}
    \end{subfigure}
    \caption{Results for the leading order $\gamma\to q\bar q $ jet EEC with exact kinematics. Jet energies are ordered from left to right and top to bottom as: $p_t=10, \, 20, \, 30, \, 50$ GeV. We show the curves for: vacuum (black), the semi-classical approximation (blue), the semi-analytic medium factor $F_{\rm med}$ obtained in the large $N_c$ limit and keeping only factorizable terms without using the semi-classical approximation detailed in the main text (red, see~\cite{Isaksen:2023nlr} for details) and the exact result extract from the numerical routine (green). We note that the error band shown is computed by comparing the semi-analytical large $N_c$ result with its full numerical counterpart, following the procedure detailed in~\cite{Isaksen:2023nlr}. Thus, the error band is only indicative of the size of discretization effects entering the numerical calculation.}
    \label{fig:Johannes_results}
\end{figure}

\begin{figure}
    \centering
    \includegraphics[width=.5\textwidth]{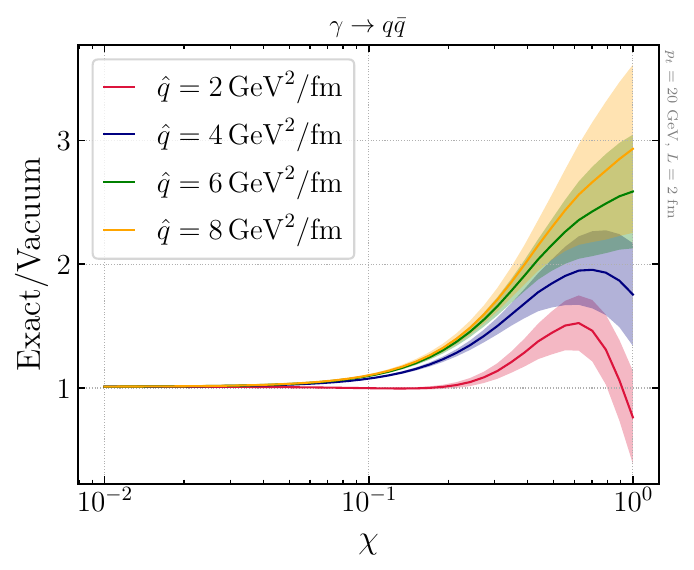}
    \caption{Medium-to-vacuum ratio of jet EEC in the $\gamma\to q\bar q $ channel using the exact matrix-element~\cite{Isaksen:2023nlr} for increasing values of $\hat q$.}
    \label{fig:Johannes_qhat_run}
\end{figure}

The previous section has clearly shown how different approximations of the in-medium matrix element can lead to disparate results at the EEC level. The absence of an exact calculation of the in-medium $q\to qg$ probability hampers any attempt to make quantitative statements utilizing analytic tools. The situation is less pessimistic in the case of a simplified scenario, namely $\gamma \to q \bar q$, thanks to a recent computation~\cite{Isaksen:2023nlr}. We use the publicly available numerical routines introduced in that paper and calculate the EEC with this exact input.~\footnote{We note that this result is still obtained in the high energy limit. Further, we do not include energy loss since it will not be necessary for the discussion and only introduces more considerable model dependence.} The result is shown in Fig.~\ref{fig:Johannes_results}, which also includes the vacuum case (black), the semi-classical approximation (blue), and a large-$N_c$ version of the exact result where only factorizable terms are kept, as discussed and computed in~\cite{Isaksen:2023nlr} (red). We note that the numerical convergence of the exact calculation degrades when probing energetic jets and/or dense media. That is why, in this study, we chose $p_t=(10,20,30,50)$ GeV and $L=2$ fm. For the same reason, the numerical error band increases as the jet energy increases. This is observed in the lowermost right panel when $\chi>0.5$. 

First, we note that despite the apparent simplicity of the $\gamma \to  q\bar q$ channel compared to the QCD one studied above, medium effects are imprinted in the EEC in a similar fashion, i.e., vacuum evolution at small angles followed by an enhancement of large-angle splittings. In addition, the exact (at $\mathcal{O}(\alpha_s)$) in-medium result shows a very similar trend to that of the naive interpolation formula discussed in the previous section. Interestingly, we observe (for the three largest energies) that the exact result (and also the large-$N_c$ curve) predicts first a depletion at relatively small angles, then a rise in the intermediate regime, and finally another depletion at large angles. We would like to highlight that the fact that the medium modified result dips twice under the vacuum line has not been observed before. In turn, the semi-classical approximation fails to grasp the exact result qualitatively and quantitatively. This reinforces the idea that any jet EEC calculation must include an accurate description of the small/large $z$ regions. We further confirmed this by computing the EEC with a higher energy suppression power, i.e., $n>1$. For the exact result, we observe that as $n$ increases, the overall distribution is suppressed, but the shape is not altered. 

 Overall, medium modifications to the splitting function seem to have a small quantitative impact on the jet EEC for this choice of medium parameters, especially as the jet energy increases. We explore other values of the jet quenching parameter $\hat q=2,4,6,8\, {\rm GeV}^2/{\rm fm}$ at fixed jet energy $p_t=20$ GeV in Fig.~\ref{fig:Johannes_qhat_run}. Increasing the density of scattering centers in the medium naturally leads to a larger enhancement. We want to remark that the medium-to-vacuum ratio remains within a factor of two, and thus, larger values of $\hat q$ do not change the qualitative picture just discussed.

Our study indicates the urge to push the accuracy of analytical calculations to assess the potential of the jet EEC to showcase medium-modifications to the splitting function or, more generally, the transition between a coherent and decoherent regime in jet evolution.

\subsection{Monte Carlo study with {\tt JetMed}}
\label{sec:results-jetmed}
To confirm the analytic findings of the previous section, we compute the EEC using a Monte Carlo approach. The interest of doing so is twofold: (i) it allows to include the effect of jet energy loss without relying on the quenching weight approximation used in Sec.~\ref{sec:e-loss}, (ii) some of the vacuum logarithmic corrections are easily accounted for, such as those coming from the running of the QCD coupling constant. The latter comment also holds for the resummation of medium-induced emissions in the regime where this effect is relevant, namely when the energy of the emitted gluons is comparable to the multiple branching scale $\omega_s$ introduced in Sec.~\ref{sec:e-loss}.

We shall use the Monte Carlo parton shower {\tt JetMed}~\cite{Caucal:2018dla,Caucal:2019uvr} whose phase space for vacuum-like radiations, as described in Sec.~\ref{sec:med_phase_space}, is already built-in and accounts for color coherence effects. In a nutshell, the {\tt JetMed} parton shower relies on the factorization in time between vacuum-like radiations and medium-induced emissions in a three-stage approach: during the first stage, the highly virtual partons produced by the hard scattering are evolved following a standard angular ordered vacuum-like shower, albeit constrained by the in-medium conditions $k_t^2\ge \sqrt{\hat{q}\omega}$ and $\theta>\theta_c$. Physically, this first cascade happens at time $t=0$ measured from the hard process; this is formally correct within the double logarithmic approximation since the in-medium condition implies that $t_f\ll L$ so that all the in-medium vacuum-like emissions happen very fast as compared to the longitudinal size of the medium. In the second step, the produced partons are subsequently evolved using an ordering in time from $t=0$ up to $t=L$ with the rate given by the BDMPS-Z rate. These medium-induced emissions also undergo transverse momentum broadening between successive branchings. Finally, in the last stage, the outgoing partons are again evolved following an angular ordered vacuum-like cascade down to the hadronization scale in the out-medium phase-space corresponding to the conditions $t_f\ge L$ or $\theta<\theta_c$. The angle of the first splitting of the out-medium shower is not constrained by its parents due to the angular ordering violation effect (anti-angular ordering) caused by the color decoherence of the parent dipole~\cite{Mehtar-Tani:2012mfa, Casalderrey-Solana:2012evi}.

In the in-medium stage, the splitting is assumed to be purely vacuum-like, meaning that the splitting probability is given by the DGLAP splitting kernel. Hence, the $F_{\rm med}$ factor introduced in Eq.~\eqref{eq:x-section_full} is not taken into account, albeit in an effective way via the phase space constraint, which is valid to double logarithmic accuracy. To test the effect of the $F_{\rm med}$ correction, we have modified the MC to include $F_{\rm med}$ for the splitting produced inside the medium. In this way, we treat $F_{\rm med}$ as a pure $\mathcal{O}(\alpha_s)$ correction to the shower instead of artificially resumming it. The functional form we use for $F_{\rm med}$ is given by Eq.\,\eqref{eq:Fmed-hard} and is therefore valid only for hard splittings with $z>\omega_c/p_t$. We emphasize that our purpose is not to faithfully account for the $F_{\rm med}$ correction but instead to quantify the relative importance of the enhancement seen at large angles in the EEC when using Eq.\,\eqref{eq:Fmed-hard} with respect to the effect of softer BDMPS-Z emissions ($z\sim \omega_s/p_t\ll \omega_c/p_t$) and energy loss on the EEC spectrum. With this in mind, it is sufficient to use a simplified version of the $F_{\rm med}$ function, which is independent of the parton flavor of the emitter, and we will therefore use Eq.\,\eqref{eq:Fmed-hard} with $\chi=\zeta=1$. The consequence of this approximation on the $q\to qg$ splitting is illustrated in Fig.\,\ref{fig:Fmed_z} (see the difference between the black and red curves).

To sample the splitting probability for the first emission inside the medium following
\begin{equation}
    \frac{\alpha_s(k_\perp)}{\pi}P_{ij}(z)\dd z\frac{\dd\theta}{\theta}\left[1+F_{\rm med}(z,\theta)\right] \, ,
\end{equation}
we use the Sudakov veto algorithm with the envelop rate
\begin{equation}
    \frac{2C_R\alpha_s(k_\perp)}{\pi}\frac{\dd k_\perp}{k_\perp}\frac{\dd\theta}{\theta}\left(1+\frac{\frac{1}{240}\hat q L^5\theta^4k_\perp^2}{1+\frac{1}{420}\hat{q}^{3/2}L^{9/2}\theta^3}\right)\, ,
\end{equation}
to generate the next splitting angle $\theta$ and transverse momentum $k_\perp$. This envelop rate accounts for the limiting behavior  of $F_{\rm med}(z,\theta)$ as $L/t_f$ goes to $0$ or infinity and presents the advantage of having a cumulative distribution function which is analytically invertible. With this envelop rate, the subsequent splitting angle $\theta$ is generated with transverse momentum $k_\perp$ and is accepted with probability given by the ratio between the exact splitting probability and its envelope. Note that since we generate $k_\perp$ and not $z$, an additional veto is imposed to ensure longitudinal momentum conservation.

In addition to the vacuum case, in which there is no quenching, we consider three physical scenarios to gauge the sensitivity of the EEC to a particular ingredient with the possible medium modifications considered in section~\ref{sec:medium}:

\begin{enumerate}
    \item a vacuum-like shower where only the first splitting is modified by the $1+F_{\rm med}$ factor. This scenario is akin to the analytic calculation in Sec.~\ref{sec:med-split-function} and in \cite{Andres:2023xwr}; 
    \item the {\tt JetMed} baseline, which does not include the $F_{\rm med}$ correction but accounts for BDMPS-Z radiations, coherent and incoherent jet energy loss, and angular ordering violation effects;
    \item the {\tt JetMed} baseline with the first vacuum-like splitting inside the medium sampled using the $1+F_{\rm med}$ correction. This scenario enables one to compare the relative impact of the $F_{\rm med}$ factor with respect to soft medium-induced emissions and large-angle energy loss.
\end{enumerate}

\begin{figure}
    \centering
    \includegraphics[width=0.49\textwidth,page=3]{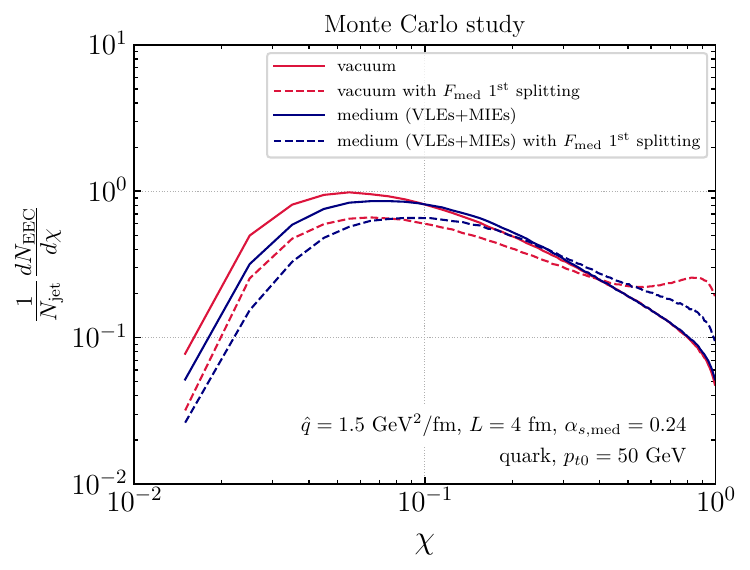}\hfill
    \includegraphics[width=0.49\textwidth,page=4]{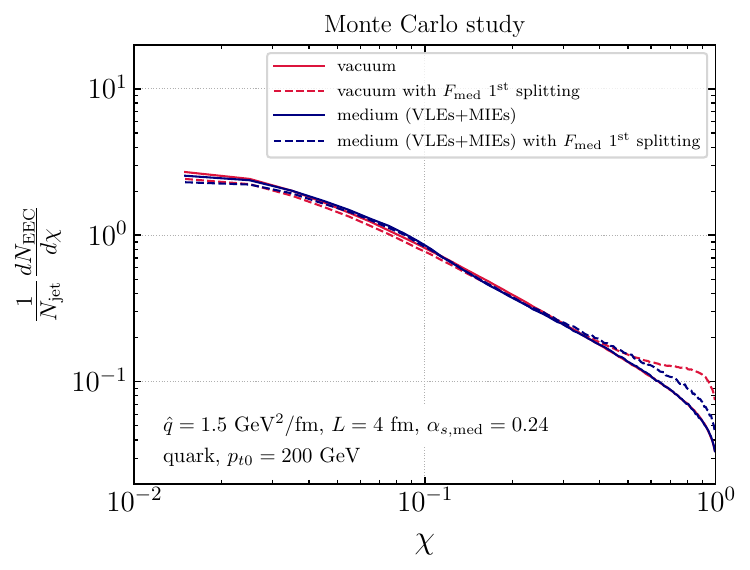}\hfill    
    \caption{Monte Carlo calculation of the EEC as a function of $\chi=\theta/R$ for a monochromatic initial hard spectrum with $p_{t0}=50$ GeV (left) and $p_{t0}=200$ GeV (right) for several medium-modified jet evolution as discussed in the main text.}
    \label{fig:JetMed1}
\end{figure}

In Fig.\,\ref{fig:JetMed1}, we show the EEC distribution in jets triggered by a quark with a fixed initial $p_{t0}=50$ GeV (left plot) and $p_{t0}=200$ GeV (right plot) in the vacuum and the medium for the three physical set-ups. Since the kinematic of the parton sourcing the jet is fixed and no initial hard spectrum is included, the jet selection has no effect, and the plot compares the EEC of the same jet population. In particular, the energy loss effect analytically implemented in Sec.~\ref{sec:e-loss} is not present in this calculation. This allows us to understand the impact on the EEC of intrinsic modifications of the jet shower in the medium. Another important point is that \texttt{JetMed} does not account for hadronization. Therefore, the curves' turnover is due to the shower cutoff and not confinement dynamics~\cite{Komiske:2022enw}. One first notices that the $F_{\rm med}$ factor computed in the semi-hard approximation is responsible for the bump observed at large $\chi$. This bump is considerably reduced when removing the $F_{\rm med}$ factor and considering only BDMPS-Z emissions, in agreement with the analytic findings of the toy model discussed towards the end of Sec.~\ref{sec:med-split-function}. The main message of this plot is that without the enhancement introduced by $F_{\rm med}$ at large angles, the intrinsic modification of the jet evolution in the medium caused by the phase space modifications and BDMPS-Z-like emissions leave almost no imprint on EEC distribution. As previously discussed, the effect of $F_{\rm med}$ is vastly overestimated since the semi-hard approximation is invalid in the kinematic regime where the enhancement appears.

\begin{figure}
    \centering
    \includegraphics[width=0.49\textwidth,page=1]{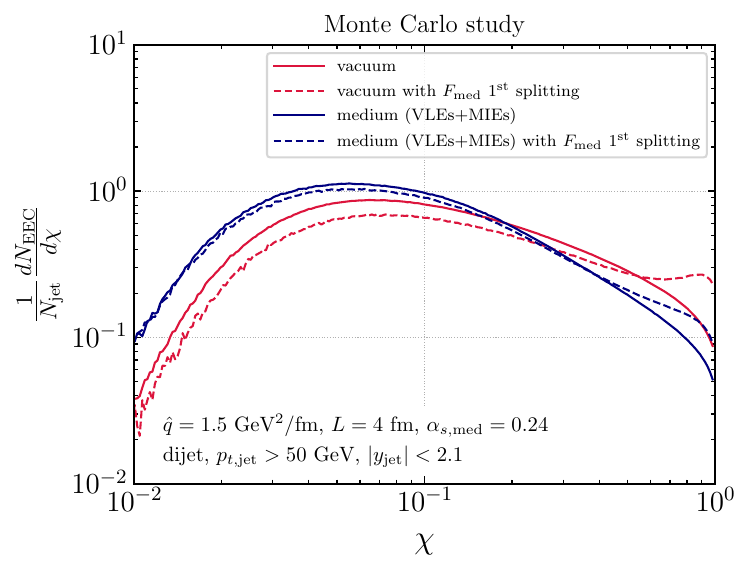}\hfill
    \includegraphics[width=0.49\textwidth,page=2]{plot-EEC-Jetmed.pdf}\hfill   
    \caption{Monte Carlo calculation of the EEC as a function of $\chi=\theta/R$ in dijet events at the LHC. Selected jets have $p_t>50$ GeV (left) or $p_t>200$ GeV (right) and absolute value of rapidity $|y|<2.1$.}
    \label{fig:JetMed2}
\end{figure}

We turn now to the calculation of EEC, including a more realistic hard scattering spectrum (generated using LO $2\to 2$ matrix elements for dijet production in $pp$ collisions at the LHC with $\sqrt{s}=5.02$ TeV) and jet selection. Introducing a jet selection on $p_t$ enables one to characterize the dependence of EECs on the energy loss effect. We expect this Monte Carlo calculation to qualitatively reproduce the analytic results obtained in Sec.~\ref{sec:e-loss} with the quenching weight method. Fig.~\ref{fig:JetMed2} shows that this is indeed the case, as one observes a suppression at large angles and an enhancement at small angles from the vacuum (solid red) to the medium (solid curve) calculation. This is a combined consequence of energy loss and jet selection since the population of jets included in the analysis differs in the vacuum and the medium. Jets selected in the medium are sourced by initial hard partons with a larger $p_{t0}$ resulting in EECs which are larger at small angles and smaller at large angles, as noticed by comparing the solid red curves between the left and right plots of Fig.\,\ref{fig:JetMed1}. After adding the modification of the first splitting in the medium in the semi-hard approximation via the $F_{\rm med}$ factor (dashed blue line), one observes a slight enhancement at large angles. There are two competing effects: energy loss at large angles via medium-induced emissions and modifications of the first splitting inside the medium, which respectively cause a suppression or an enhancement at large angles in the distribution. Given that the functional form of $F_{\rm med}$ implemented in the Monte Carlo should be considered as an upper limit of a more realistic value, one can safely conjecture that the dominant effect among these two competing mechanisms will be the energy loss.

\section{Lund-based definition of the EEC}
\label{sec:LEEC}

Throughout this paper, we have studied the canonical definition of the EEC in which the energy weight is set to $1$, i.e., $n=1$ in Eq.~\eqref{eq:eec-definition}. In a heavy-ion context, higher values of $n$ might be helpful to, for example, mitigate the overwhelming underlying event. However, setting $n>1$ leads to a collinear unsafe observable. For $n=2$, it has been shown that these divergences can be absorbed into moments of non-perturbative objects such as fragmentation/track functions in proton-proton collisions~\cite{Chen:2020vvp,Li:2021zcf}. Extending this approach to heavy-ion collisions is a challenging task for multiple reasons, the main one being the multi-scale nature of the problem~\cite{tracks_paper}.\footnote{Appendix B of Ref.~\cite{Andres:2023xwr} showed results for the jet EEC with $n=2$ neglecting non-perturbative ingredients.} It would be ideal to minimally modify the definition of the EEC such that higher powers of the energy weight can be accommodated without sacrificing perturbative calculability. 

We do so in the Lund family of observables context~\cite{Dreyer:2018nbf} and denote this alternative definition as Lund-based EEC or LEEC. The algorithmic procedure to calculate the Lund-based definition of the energy-energy correlator goes as follows. The starting point is to recluster a given jet (typically defined with the anti-$k_t$ algorithm~\cite{Cacciari:2008gp}) with Cambridge/Aachen~\cite{Dokshitzer:1997in,Wobisch:1998wt} to obtain an angular-ordered sequence. Then, we follow the traditional recipe in Lund-based observables:
\begin{itemize}
\item Undo the last-clustering step to generate two subjets, $j_1$ and $j_2$.
\item Calculate their relative $k_t$ defined as $k_t=\min(x_1,x_2)\Delta R_{12}$, where the concrete definitions of $x$ (an energy-like variable) and $\Delta R_{12}$  (an angular-like variable) depend on the collision system. For $e^+e^-$, $x_i=E_i$ and $\Delta=\theta_{ij}$, while in $pp$ in $x_i=p_{ti}$ and $\Delta R_{ij} = \sqrt{(y_i-y_j)^2 + (\phi_i-\phi_j)^2}$. 
\item Only when $k_t>k_{t,\rm cut}$, record the softest branch, so-called primary Lund declustering.
\item Repeat from step 1 following only the hardest subjet, i.e., the primary branch. 
\item Once there is nothing left to decluster, calculate the EEC as
\begin{equation}
\label{eq:leec}
\frac{\dd \Sigma^{(n)}}{\dd \chi} = \frac{1}{\sigma}\sum_{ \{i,j\}\in {\rm declust.} }\int_0^1 \dd z \frac{\dd\sigma}{\dd \theta_{ij} \dd z} z^n(1-z)^n \delta\left(\chi-\frac{\theta_{ij}}{R}\right)\Theta(k_t>k_{t,\rm cut})\, ,
\end{equation}
where the sum runs over all primary Lund declusterings.
\end{itemize}

The main advantage of using subjets is that Eq.~\eqref{eq:leec} remains collinear safe for any value of $n$. In the following exploratory study, we will disregard the $k_t>k_{t,\rm cut}$ condition and present some of its properties using Monte Carlo simulations. The logarithmic resummation of this new observable will be the subject of a separate publication~\cite{leec_paper}.

\begin{figure}
    \centering
    \includegraphics[width=0.49\textwidth]{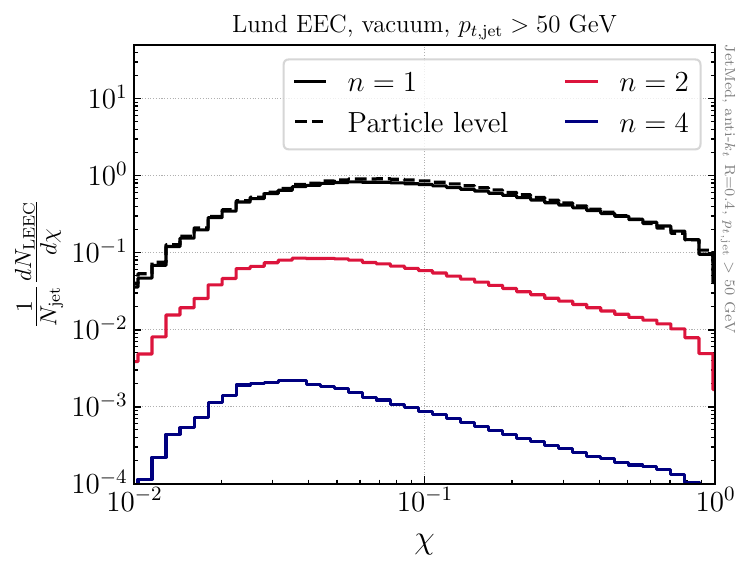}\hfill
    \includegraphics[width=0.49\textwidth]{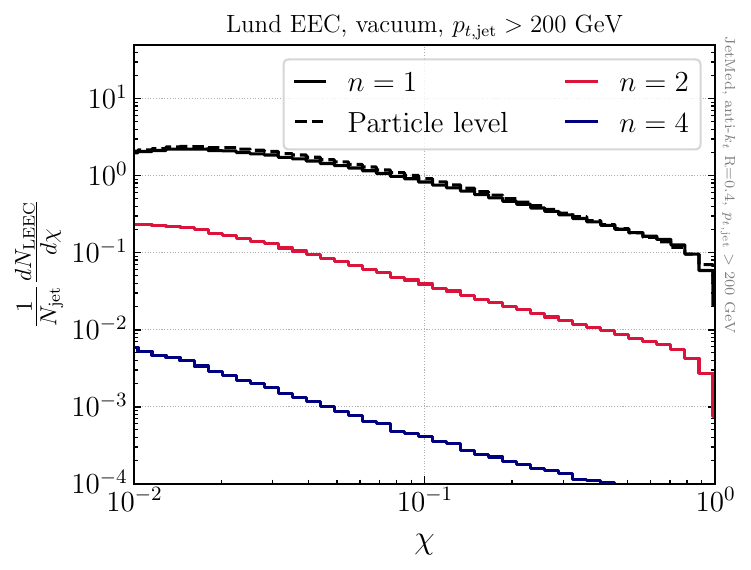}\hfill    
    \caption{Lund-based EEC for several values of the energy weighting parameter $n$ in proton-proton collisions for two jet $p_t$ selections: $p_t>50$ GeV (left) and $p_t>200$ GeV (right). Note that for $n=1$ the particle level result is indicated by a black, dashed line.}
    \label{fig:LEEC1}
\end{figure}

In Fig.~\ref{fig:LEEC1}, we present results for the Lund EEC with $k_{t,\rm cut}=0$ GeV,in proton-proton collisions using the \texttt{JetMed} samples from the previous section. For $n=1$ we plot both the particle level (i.e., the traditional EEC) and the Lund EEC. We observe very small differences between the two cases. When increasing $n$, we only show results for the Lund version. The overall suppression is naturally explained as a reduction in the contribution of each pair of subjets to the observable due to the energy-weight penalty. Interestingly, the slope of the EEC at large angles becomes steeper with increasing $n$. This might be related to different anomalous dimensions entering into the logarithmic structure although no direct link can be made from the parton shower result unless taking the appropriate limits, as introduced in Ref.~\cite{Dasgupta:2020fwr}. Regarding the $p_t$-dependence, we observe the expected overall shift to smaller angles of the full distribution and quantitatively small differences between the Lund approach and the standard definition for $n=1$.

\begin{figure}
    \centering
    \includegraphics[width=0.49\textwidth,page=1]{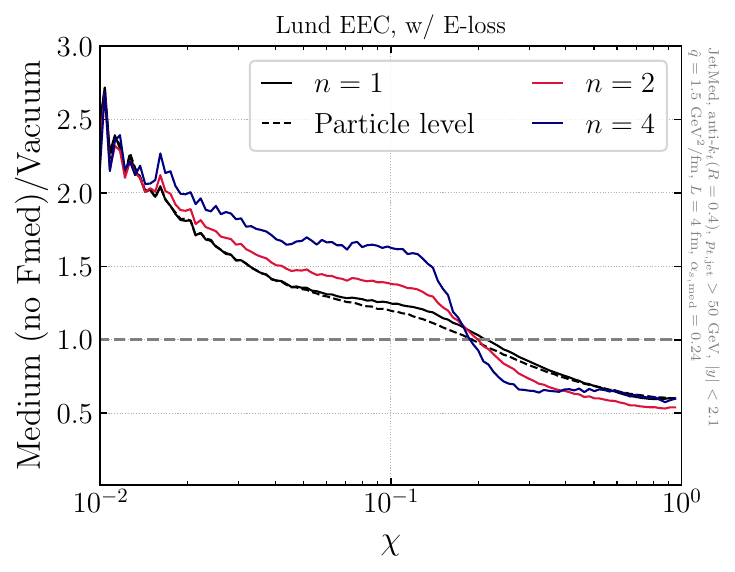}\hfill
    \includegraphics[width=0.49\textwidth,page=2]{plot-LEEC-ndep.pdf}\hfill 
    \caption{Medium-to-vacuum ratio of the Lund-based EEC for several values of the energy weighting parameter $n$ with (right) or without (left) $F_{\rm med}$ correction to the first splitting.}
    \label{fig:LEEC2}
\end{figure}

The impact of medium corrections to the Lund EEC (with $k_{t,\rm cut}=0$ GeV) is explored in Fig.~\ref{fig:LEEC2}. The left panel does not include $F_{\rm med}$, while the right one does. In both cases, energy loss is part of the simulation. As in vacuum, the Lund-based result for $n=1$ is almost identical to the standard definition of the EEC. Larger values of $n$ reveal a larger medium-to-vacuum ratio in the entire angular range with or without $F_{\rm med}$. To explain this behavior, let us first focus on the case where $F_{\rm med}$ is switched off, and the Lund EEC exhibits an apparent narrowing. The net effect of increasing $n$ is to reduce the number of pairs that give a sizeable contribution to the EEC, i.e., it should become dominated by a handful of hard subjet pairs the larger $n$ is. In other words, the EEC resembles a SoftDrop-like~\cite{Larkoski:2014wba} observable such as $\theta_g$. We note that the pronounced narrowing of the $n=4$ EEC result is quantitatively similar to that of $\theta_g$~\cite{ALargeIonColliderExperiment:2021mqf,Caucal:2021cfb,Caucal:2020uic}. It would be interesting to simultaneously calculate/measure these two observables and understand whether they share the same transition point. Including $F_{\rm med}$ compensates for this narrowing and leads to the characteristic enhancement around the jet boundary, as observed in the right panel of Fig.~\ref{fig:LEEC2}. 

\section{Conclusions}
\label{sec:conclusions}
The study of parton shower evolution and hadronization in terms of energy-energy correlators has gained momentum in the last couple of years, theoretically and experimentally. While high-precision calculations exist for the vacuum baseline~\cite{Duhr:2022yyp,Chen:2023zlx}, much less is known about the formal accuracy of EEC calculations in the heavy-ion context. In this work, we have provided a detailed study of the EEC in a dense QCD background, both with logarithmic resummation tools and Monte Carlo simulations. In summary, our results demonstrate that certain theoretical approximations to the in-medium matrix element previously employed in the literature are insufficient to grasp the dynamics of this observable fully and can lead to a significant overestimate of medium modifications. We also show that a leading-order description of the EEC can be significantly spoiled by other well-known jet-quenching effects, such as energy loss.

On the analytic front, we have revisited the LL resummation of the jet EEC following a diagrammatic approach that, in our view, is more flexible than other approaches when it comes to including medium effects. We hope this will allow for higher-order calculations taking into account medium effects; we leave related efforts to future work. Another key result of this work is the first full leading order EEC calculation in the medium for the $\gamma \to q\bar q$ channel, which shares many features with pure QCD channels. The exact results confirm that commonly used semi-analytic approximations vastly overestimate the medium enhancement. We hope this result will trigger new developments in the computation of the exact $q\to qg$ in-medium matrix element. 

Monte Carlo simulations of the EEC further support these conclusions. Our calculation includes part of the NLL resummation in vacuum (such as running coupling corrections), energy loss via medium-induced emissions, and medium modification of the first in-medium splitting.  We find that the dominant effect is caused by energy loss, which suppresses the medium distribution at large angles and increases it at small angles (with the transition occurring around $\theta_c$). Even when overestimating the medium modification to the first in-medium semi-hard splitting, we observe that the energy loss effect can overwhelm the large angle enhancement. Further steps to improve our Monte Carlo calculation include (i) to have an NLL accurate parton shower (see \cite{Dasgupta:2020fwr,vanBeekveld:2022zhl,vanBeekveld:2022ukn,vanBeekveld:2023lfu}) for the vacuum-like evolution, (ii) to perform a proper matching using the exact $F_{\rm med}$ factor at the level of the short distance cross-section, to have control over the $\mathcal{O}(\alpha_s)$ $1\to 2$ in-medium matrix element within a parton shower approach. 

Finally, we argue that the energy weight in the EEC definition can be used as a knob to amplify medium effects. To not spoil collinear safety, we introduce an extension of the standard jet EEC definition, in which the building blocks are primary Lund declusterings instead of individual particles. We anticipate that this gain in terms of sensitivity to medium effects might come at the price of a more complex all-order structure. Nevertheless, the resummation of this new observable at NLL accuracy can be achieved numerically within the PanScales framework and will be presented in a separate publication.

\section*{Acknowledgments}
The Feynman diagrams in this work were produced using the {\tt JaxoDraw} software~\cite{Binosi:2003yf}. JB and RS are supported by the United States Department of Energy under Grant Contract DESC0012704. We express our heartfelt gratitude to Pier Monni for insightful discussions and collaboration on related topics. We are grateful to the authors of Ref.~\cite{Andres:2022ovj} for a careful reading of the manuscript and helpful suggestions. We thank Laura Havener, Peter Jacobs, Mateusz Ploskon, and Wenqing Fan for discussions on the experimental measurement of the EEC in heavy ions. We are thankful to Yacine Mehtar-Tani, Guilherme Milhano, and Xin-Nian Wang for many helpful discussions. JB is also grateful to Ian Moult and Andrey Sadofyev for useful parallel discussions, and Swagato Mukherjee for sharing interesting related ideas. PC would like to thank Feng Yuan for his hospitality at LBNL during the completion of this project. JB would like to thank the hospitality of the CERN TH division during the completion of this project.

\bibliographystyle{elsarticle-num}
\bibliography{references.bib}

\end{document}